%% file: dieloesung.tex
\documentclass[twocolumn,epjc3]{svjour3}  

\journalname{Eur. Phys. J. C}

\smartqed  

\RequirePackage{fix-cm}
\RequirePackage{graphicx}
\RequirePackage{mathptmx}      

\RequirePackage{hyperref}
\RequirePackage{xcolor}
\RequirePackage{amsmath}
\RequirePackage{amssymb,wasysym}
\usepackage{cite}
\usepackage{booktabs,ulem}
\usepackage{threeparttable}
\usepackage{multirow}
\usepackage[hang,raggedright]{subfigure}

\newcounter{setcounter}

\newcommand{\bsmm}{\mathcal{B}(B_s\to\mu\mu)}
\newcommand{\bsgamma}{\mathcal{B}(b\to s\gamma)}
\newcommand{\btaunu}{\mathcal{B}(B\to\tau\nu)}

\graphicspath{{figures/}}

\begin{document}

\title{Killing the cMSSM softly
}

\subtitle{}


\author{
  Philip~Bechtle\thanksref{email:bechtle,addr:bonn}
  \and Jos\'e~Eliel~Camargo-Molina\thanksref{email:camargo,addr:lund}
  \and Klaus~Desch\thanksref{email:desch,addr:bonn}
  \and Herbert~K.~Dreiner\thanksref{email:dreiner,addr:bonn,addr:bethe}
  \and Matthias~Hamer\thanksref{email:hamer,addr:rio}
  \and Michael~Kr\"amer\thanksref{email:kraemer,addr:aachen}
  \and Ben~O'Leary\thanksref{email:oleary, addr:wuerzburg}
  \and Werner~Porod\thanksref{email:porod, addr:wuerzburg}
  \and Bj\"orn~Sarrazin\thanksref{email:sarrazin,addr:bonn}
  \and Tim~Stefaniak\thanksref{email:stefaniak, addr:ucsc}
  \and Mathias~Uhlenbrock\thanksref{email:uhlenbrock,addr:bonn}
  \and Peter~Wienemann\thanksref{email:wienemann,addr:bonn}
}


\thankstext{email:bechtle}{e-mail: bechtle@physik.uni-bonn.de}
\thankstext{email:camargo}{e-mail: jose.camargo@physik.uni-wuerzburg.de}
\thankstext{email:desch}{e-mail: desch@physik.uni-bonn.de}
\thankstext{email:dreiner}{e-mail: dreiner@uni-bonn.de}
\thankstext{email:hamer}{e-mail: mhamer@cbpf.br}
\thankstext{email:kraemer}{e-mail: mkraemer@physik.rwth-aachen.de}
\thankstext{email:oleary}{e-mail: ben.oleary@physik.uni-wuerzburg.de}
\thankstext{email:porod}{e-mail: porod@physik.uni-wuerzburg.de}
\thankstext{email:sarrazin}{e-mail: sarrazin@physik.uni-bonn.de}
\thankstext{email:stefaniak}{e-mail: tistefan@ucsc.edu}
\thankstext{email:uhlenbrock}{e-mail: uhlenbrock@physik.uni-bonn.de}
\thankstext{email:wienemann}{e-mail: wienemann@physik.uni-bonn.de}


\institute{Physikalisches Institut, University of Bonn, Germany \label{addr:bonn}
  \and Department of Astronomy and Theoretical Physics,
Lund University, SE 223-62 Lund, Sweden \label{addr:lund}
  \and Bethe Center for Theoretical Physics, University of Bonn, Germany \label{addr:bethe}
  \and Centro Brasileiro de Pesquisas Fisicas, Rio de Janeiro, Brazil \label{addr:rio}
  \and Institute for Theoretical Particle Physics and Cosmology, RWTH Aachen, Germany \label{addr:aachen}
  \and Institut f\"ur Theoretische Physik und Astrophysik, University of W\"urzburg, Germany \label{addr:wuerzburg}
  \and Santa Cruz Institute for Particle Physics, University of
  California, Santa Cruz, CA 95064, USA \label{addr:ucsc}  
}

\date{\today}

\maketitle

\abstract{We investigate the constrained Minimal Supersymmetric
  Standard Model (cMSSM) in the light of constraining experimental and
  observational data from precision measurements, astrophysics, direct
  supersymmetry searches at the LHC and measurements of the properties
  of the Higgs boson, by means of a global fit using the program
  \textsc{Fittino}.  As in previous studies, we find rather poor
  agreement of the best fit point with the global data. We also
  investigate the stability of the electro-weak vacuum in the
  preferred region of parameter space around the best fit point.  We
  find that the vacuum is metastable, with a lifetime significantly
  longer than the age of the Universe.  For the first time in a global
  fit of supersymmetry, we employ a consistent methodology to evaluate
  the goodness-of-fit of the cMSSM in a frequentist approach by
  deriving $p$-values from large sets of toy experiments. We analyse
  analytically and quantitatively the impact of the choice of the
  observable set on the $p$-value, and in particular its dilution when
  confronting the model with a large number of barely constraining
  measurements.  Finally, for the preferred sets of observables, we
  obtain $p$-values for the cMSSM below 10\%, \textit{i.e.} we exclude
  the cMSSM as a model at the 90\% confidence level.  }

\input{introduction}

\input{methods}
\input{observables}

\input{results}

\input{conclusions}

\input{acknowledgements}


\input{dieloesung.bbl}
\end{document}

%% file: introduction.tex
\section{Introduction}
\label{sec:Introduction}

Supersymmetric theories~\cite{Wess:1974tw,Golfand:1971iw} offer a unique extension of the external symmetries of the Standard Model (SM) with spinorial generators~\cite{Haag:1974qh}. Due to the experimental constraints on the supersymmetric masses, supersymmetry must be broken.  Supersymmetry allows for the unification of the electromagnetic, weak and strong gauge couplings~\cite{Langacker:1991an,Amaldi:1991cn,Ellis:1990wk}.  Through radiative symmetry breaking~\cite{Nilles:1982dy,Ibanez:1982fr}, it allows for a dynamical connection between supersymmetry breaking and the breaking of SU(2)$\times$U(1), and thus a connection between the unification scale and the electroweak scale. Furthermore, supersymmetry provides a solution to the fine--tuning problem of the SM~\cite{Gildener:1976ai,Veltman:1980mj}, if at least some of the supersymmetric particles have masses below or near the TeV scale~\cite{Barbieri:1987fn}. Furthermore, in supersymmetric models with $R$-parity conservation~\cite{Fayet:1977yc,Farrar:1978xj}, the lightest supersymmetric particle (LSP) is a promising candidate for the dark matter in the universe~\cite{Goldberg:1983nd,Ellis:1983ew}.

Of all the implementations of supersymmetry, there is one which has stood out throughout in phenomenological and experimental studies: The constrained Minimal Supersymmetric Standard Model (cMSSM) \cite{Nilles:1983ge,Martin:1997ns}. As we will show in this paper, albeit it is a simple model with a great set of benefits over the SM, it has come under severe experimental pressure. To explain and -- for the first time -- to quantify this pressure is the aim of this paper.

The earliest phenomenological work on supersymmetry was performed almost 40 years ago~\cite{Farrar:1978xj,Fayet:1977yc,Fayet:1974pd,Fayet:1974jb,Fayet:1976et} in the framework of global supersymmetry.  Due to the mass sum rule~\cite{Ferrara:1979wa}, realistic models require local supersymmetry, or supergravity~\cite{Freedman:1976xh,Cremmer:1978hn,Cremmer:1982en,Nilles:1983ge}.  The cMSSM is an effective parametrisation motivated by realistic supergravity models. Since we wish to critically investigate the viability of the cMSSM in detail here, it is maybe in order to briefly recount some of its history.

The cMSSM as we know it was first formulated in~\cite{Kane:1993td}. However, it is based on a longer development in the construction of realistic supergravity models.  A globally supersymmetric model with explicit soft supersymmetry breaking~\cite{Girardello:1981wz} added by hand was first introduced in~\cite{Dimopoulos:1981zb}. It is formulated as an SU(5) gauge theory, but is otherwise already very similar to the cMSSM, as we study it at colliders. It was however not motivated by a fundamental supergravity theory. A first attempt at a realistic model of spontaneously breaking local supersymmetry and communicating it with gravity mediation is given in~\cite{Nilles:1982ik}. At tree-level, it included only the soft breaking gaugino masses. The soft scalar masses were generated radiatively. The soft breaking masses for the scalars were first included in~\cite{Barbieri:1982eh,Chamseddine:1982jx}. Here both the gauge symmetry and supersymmetry are broken spontaneously~\cite{Cremmer:1982en}. In~\cite{Chamseddine:1982jx} the first locally supersymmetric grand unified model was constructed.  Connecting the breaking of SU(2)$\times$U(1) to supersymmetry breaking was first presented in~\cite{Nilles:1982dy}, this included for the first time the bi- and trilinear soft-breaking $B$ and $A$ terms.  Radiative electroweak symmetry breaking was given in~\cite{Ibanez:1982fr}. A systematic presentation of the low--energy effects of the spontaneous breaking of local supersymmetry, which is communicated to the observable sector via gravity mediation is given in~\cite{Soni:1983rm,Hall:1983iz}.

Thus all the ingredients of the cMSSM, the five parameters $M_0,\,M_{1/2},\,\tan\beta,\,\mathrm{sgn}(\mu), \,A_0$ were present and understood in early 1982. Here $M_0$ and $M_{1/2}$ are the common scalar and gaugino masses, respectively, and $A_0$ is a common trilinear coupling, all defined at the grand unified scale. The ratio of the two Higgs vacuum expectation values is denoted by $\tan\beta$, and $\mu$ is the superpotential Higgs mass parameter. Depending on the model of supersymmetry breaking there were various relations between these parameters. By the time of~\cite{Kane:1993td}, no obvious simple model of supersymmetry breaking had been found, and it was more appropriate to parametrise the possibilities for phenomenological studies, in terms of these five parameters.  In many papers the minimal supergravity model (mSUGRA) is often deemed synonymous with the cMSSM.  However, more precisely mSUGRA contains an additional relation between $A_0$ and $M_0$ reducing the number of parameters~\cite{AbdusSalam:2011fc}.

The cMSSM is a very well-motivated, realistic and concise supersymmetric extension of the SM. Despite the small number of parameters, it can incorporate a wide range of phenomena. To find or to exclude this model has been the major quest for the majority of the experimental and phenomenological community working on supersymmetry over the last 25 years.

In a series of \textsc{Fittino} analyses~\cite{Bechtle:2009ty,Bechtle:2011dm,Bechtle:2012zk,Bechtle:2013mda} we have confronted the cMSSM to precision observables, including in particular the anomalous magnetic moment of the muon, $(g-2)_\mu$, astrophysical observations like the direct dark matter detection bounds and the dark matter relic density, and collider constraints, in particular from the LHC experiments, including the searches for supersymmetric particles and the mass of the Higgs boson.

Amongst the previous work on understanding the cMSSM in terms of global analyses, there are both those applying frequentist statistics~\cite{Buchmueller:2010ai,Buchmueller:2011aa,Buchmueller:2011ki,Buchmueller:2011sw, Buchmueller:2011ab,Buchmueller:2012hv,Buchmueller:2013psa,Buchmueller:2013rsa, Buchmueller:2014yva,deVries:2015hva,Bagnaschi:2015eha,Ellis:2013oxa,Bertone:2011nj,Strege:2011pk,Strege:2012bt, Balazs:2013qva,Beskidt:2012bh,Beskidt:2014oea,Lafaye:2007vs,Adam:2010uz,Henrot-Versille:2013yma} and Bayesian statistics~\cite{Allanach:2007qk,Allanach:2011ut,Allanach:2011wi,deAustri:2006pe,Fowlie:2011mb,Roszkowski:2012uf,Fowlie:2012im,Kowalska:2012gs, Kowalska:2013hha,Kowalska:2014hza,Roszkowski:2014wqa,Kowalska:2015kaa}. While the exact positions of the minima depend on the statistical interpretation, they agree on the overall scale of the preferred parameter region.

  We found that the cMSSM does not provide a good description of all observables. In particular, 
our best fit predicted supersymmetric particle masses in the TeV range or above, \textit{i.e.}\ possibly beyond the 
reach of current and future LHC searches. The precision observables like $(g-2)_\mu$ or the branching ratio of 
$B$ meson decay into muons, BR$(B_s\to \mu\mu)$, were predicted very close to their SM value, and no signal
for dark matter in direct and indirect searches was expected in experiments conducted at present or in the near 
future. 

According to our analyses, the Higgs sector in the cMSSM consists of a light scalar Higgs boson with SM-like 
properties, and heavy scalar, pseudoscalar and charged Higgs bosons beyond the reach of current and future 
LHC searches. We also found that the LHC limits on supersymmetry and the large value of the light scalar Higgs 
mass drives the cMSSM into a region of parameter space with large fine tuning. See also~\cite{Cassel:2011tg,
Ellwanger:2011mu,Kaminska:2013mya,Ghilencea:2013nxa,Baer:2014ica} on fine-tuning. We thus concluded that 
the cMSSM has become rather unattractive and dull, providing a bad description of experimental observables like 
$(g-2)_\mu$ and predicting grim prospects for a discovery of supersymmetric particles in the near future~\cite{Bechtle:2014yna}.

While our conclusions so far were based on a poor agreement of the best fit points with data, as expressed in a rather high ratio of the global $\chi^2$ to the number of degrees of freedom, there has been no successful quantitative evaluation of the "poor agreement" in terms of a \textit{confidence level}. Thus, the cMSSM could not be \textit{excluded} in terms of frequentist statistics due to the lack of appropriate methods or the numerical applicability.

Traditionally, a hypothesis test between two alternative hypotheses, based on a likelihood ratio, would be employed 
for such a task. An example for this is \textit{e.g.}~the search for the Higgs boson, where the SM without a 
kinematically accessible Higgs as a ``null hypothesis" is compared to an alternative hypothesis of a SM with a given 
accessible Higgs Boson mass. However, in the case employed here, there is a significant problem with this approach: 
The SM does not have a dark matter candidate and thus is highly penalised by the observed cold dark matter content 
in the universe. (It is actually excluded.) Thus, the likelihood ratio test will \textit{always} prefer the supersymmetric 
model \textit{with} dark matter against the SM, no matter how bad the actual goodness-of-fit might be.

Thus, in the absence of a viable null hypothesis without
supersymmetry, in this paper we address this question by calculating
the $p$-value from repeated fits to randomly generated
pseudo-measurements. The idea to do this has existed before (see \textit{e.g.}~\cite{Fowlie:Thesis}), but due to the very high demand in CPU power,
specific techniques for the re-interpretation of the parameter scan
had to be developed to make such a result feasible for the first time. 
 In addition to the previously employed observables, here we included the 
measured Higgs boson signal strengths in detail. We find that the observed $p$-value depends 
sensitively on the precise choice of the set of observables.

The calculation of a $p$-value allows us to quantitatively address the question, whether a \textit{non-trivial} cMSSM can be distinguished from a cMSSM which, due to the decoupling nature of SUSY, effectively resembles the SM plus generic dark matter.

The paper is organised as follows. In Sec.~\ref{sec:methods} we describe the method of determining the $p$-value from pseudo measurements. The set of experimental observables included in the fit is presented in Sec.~\ref{sec:observables}. The results of various fits with different sets of Higgs observables are discussed in Sec.~\ref{sec:results}. Amongst the results presented here are also predictions for direct detection experiments of dark matter, and a first study of the vacuum stability of the cMSSM in the full area preferred by the global fit.
We conclude in
Sec.~\ref{sec:conclusions}.



%% file: methods.tex
\section{Methods}\label{sec:methods}

In this section, we describe the statistical methods employed in the fit. These include the scan of the 
parameter space, as well as the determination of the $p$-value.  Both are non-trivial, because of the 
need for ${\cal O}(10^9)$ theoretically valid scan points in the cMSSM 
parameter space, where each 
point uses about 10 to 20 seconds of CPU time.  Therefore, in this paper optimised scanning 
techniques are used, and a technique to re-interpret existing scans in pseudo experiments (or 
``toy studies'') is developed specifically for the task of determining the frequentist $p$-value of a SUSY 
model for the first time.

\subsection{Performing and Interpreting the Scan of the Parameter Space}

In this section, the specific Markov Chain Monte Carlo (MCMC) method used in the scan, the 
figure-of-merit used for the sampling, and the (re-)interpretation of the cMSSM parameter points 
in the scan is explained. 

\subsubsection{Markov Chain Monte Carlo method }
The parameter space is sampled using a MCMC method based on 
the Metropolis-Hastings algorithm~\cite{SIC:637496,Metropolis:1953am,Hastings:1970aa}. 
At each tested point in the parameter space the model predictions for all observables are calculated 
and compared to the measurements. The level of agreement between predictions and measurements 
is quantified by means of a total $\chi^2$, which in this case corresponds to the ``Large Set'' of 
observables introduced in Section~\ref{sec:observables:measurements}\footnote{Since the allowed 
region for all observable sets tested in Section~\ref{sec:results} differ only marginally, it does not matter 
significantly which observable set is chosen for the initial scan, as long as it efficiently samples
the relevant parameter space}:
\begin{eqnarray}
\chi^{2} = \left( \vec{O}_{\textrm{meas}} - \vec{O}_{\textrm{pred}}\right )^{T}\textrm{cov}^{-1} \left( \vec{O}
_{\textrm{meas}} - \vec{O}_{\textrm{pred}} \right ) + \chi^{2}_{\textrm{limits}},
\end{eqnarray}
where $\vec{O}_{\textrm{meas}}$ is the vector of measurements, $\vec{O}_{\textrm{pred}}$ the 
corresponding vector of predictions, $\textrm{cov}$ the covariance matrix including theoretical 
uncertainties and $\chi^{2}_{\textrm{limits}}$ the sum of all $\chi^{2}$ contributions from the
employed limits, \textit{i.e.} the quantities for which bounds, but no measurements are applied.

After the calculation of the total $\chi^{2}$ at the $n^{th}$ point in the Markov chain, a new point is determined by 
throwing a random number according to a probability density called proposal density. We use Gaussian proposal densities, where 
for each parameter the mean is given by the current parameter value and the width is regularly adjusted 
as discussed below.

The $\chi^{2}$ for the $(n+1)^{th}$ point is then calculated and compared to the $\chi^{2}$ for the ${n}^{th}$ point. If the new point shows better or equal agreement between predictions and measurements, 
\begin{eqnarray}
\chi^{2}_{n+1} \leq \chi^{2}_{n},
\end{eqnarray}
it is accepted.
If the $(n+1)^{th}$ point shows worse agreement between the predictions and measurements, it is accepted with probability 
\begin{eqnarray}
\varrho = \textrm{exp}\left( - \frac{\chi^{2}_{n+1} - \chi^{2}_{n}}{2} \right ), 
\end{eqnarray}
and rejected with probability $1-\varrho$. If the $(n+1)^{th}$ point is rejected, new parameter values are generated 
based on the $n^{th}$ point again. If the $(n+1)^{th}$ point is accepted, new parameter values are generated 
based on the \mbox{$(n+1)^{th}$} point.  Since the primary goal of using the MCMC method is the accurate 
determination of the best fit point and a high sampling density around that point in the region of $\Delta\chi^2\leq6$, 
while allowing the MCMC method to escape from local minima in the $\chi^{2}$ landscape, it is mandatory to neglect 
rejected points in the progression of the Markov chain.  
However, the rejected points may well be used in the frequentist interpretation of the Markov chain and for the determination of the $p$-value. Thus, we store them as well in order to increase the overall sampling density.

An automated optimisation procedure was employed to determine the width of the Gaussian proposal densities 
for each parameter for different targets of the acceptance rate of proposed points. Since the frequentist 
interpretation of the Markov chain does not make direct use of the point density, we can employ chains, where the 
proposal densities vary during their evolution and in different regions of the parameter space. We update the widths 
of the proposal densities based on the variance of the last ${\cal O}(500)$ accepted points in the Markov chain.  Also, 
different ratios of proposal densities to the variance of accepted points are used for chains started in different 
parts of the parameter space, to optimally scan the widely different topologies of the $\chi^2$ surface at different 
SUSY mass scales. These differences stem from the varying degree of correlations between different parameters 
required to stay in agreement with the data, and from non-linearities between the parameters and observables. 
They are also the main reason for the excessive amount of points needed for a typical SUSY scan, as 
compared to more nicely behaved parameter spaces. It has been ensured that a sufficient number of statistically 
independent chains yield similar scan results over the full parameter space. For the final interpretation, all 
statistically independent chains are added together.

A total of 850 million valid points have been tested. The point with the lowest overall $\chi^{2} = \chi^{2}_{\textrm{min}}$ is identified as the best fit point. 

\subsubsection{Interpretation of Markov chain results}
In addition to the determination of the best fit point it is also of interest to set limits in the cMSSM parameter space. For the Frequentist interpretation the measure
\begin{eqnarray}
\Delta \chi^{2} = \chi^{2} - \chi^{2}_{\textrm{min}}
\end{eqnarray}
is used to determine the regions of the parameter space which are excluded at various confidence levels. For this 
study the one dimensional $1\sigma$ region ($\Delta \chi^{2} < 1$) and the two dimensional $2\sigma$ region 
($\Delta \chi^{2} < 6$) are used. In a Gaussian model, where all observables depend linearly on all 
parameters and where all uncertainties are Gaussian, this would correspond to the 1-dimensional 68\% and 
2-dimensional 95\% Confidence Level (CL) regions. The level of observed deviation from this pure Gaussian approximation shall be 
discussed together with the results of the toy fits, which are an ideal tool to resolve these differences.

\subsection{Determining the $p$-value}\label{sec:methods:pvalue}
In all previous instances of SUSY fits, no true frequentist $p$-value for the fit is calculated. Instead, usually the $\chi^{2}_{min}/$ndf is 
calculated, from which for a linear model with Gaussian observables a $p$-value can easily derived. It has been 
observed that the $\chi^2_{min}/$ndf of constrained SUSY model fits such as the cMSSM have been degrading 
while the direct limits on the sparticle mass scales from the LHC got stronger (see e.g.~\cite{Bechtle:2009ty,Bechtle:2011dm,Bechtle:2012zk}). Thus, there is the 
widespread opinion that the cMSSM is obsolete. However, as the cMSSM is a highly non-linear model and the observable set 
includes non-Gaussian observables, such as one-sided limits and the ATLAS 0-lepton search, it is not obvious that the Gaussian $\chi^{2}$-distribution 
for ndf degrees of freedom can be used to calculate an accurate $p$-value for this model. Hence the main question in this 
paper is: \textit{How} obsolete is the cMSSM, exactly? To answer this, a machinery to re-interpret the scan described 
above had to be developed, since re-scanning the parameter space for each individual toy observable set is 
computationally prohibitive at present. Because during this re-interpretation of the original scan a multitude of 
different cMSSM points might be chosen as optima of the toy fits, such a procedure sets high demands on 
the scan density also over the entire approximate 2 to 3 sigma region around the observed optimum.

\subsubsection{General Procedure}
After determining the parameter values that provide the best combined description of the observables suitable 
to constrain the model, the question of the $p$-value for that model remains: Under the assumption 
that the tested model at the best fit point is the true description of nature, what is the probability $p$ to get a 
minimum $\chi^{2}$ as bad as, or worse than, the actual minimum $\chi^2$?

For a set of observables with Gaussian uncertainties, this probability is calculated by means of the $\chi^{2}$-
distribution and is given by the regularised Gamma function, $p = P\left ( \frac{n}{2}, \frac{\chi^{2}_{\textrm{min}}}
{2} \right )$. Here, $n$ is the number of degrees of freedom of the fit, which equals the number of observables 
minus the number of free parameters of the model.

In some cases, however, this function does not describe the true distribution of the $\chi^{2}$. Reasons for a 
deviation include non-linear relations between parameters and observables 
(as evident in the cMSSM, where a strong variation of the observables with the parameters at low parameter 
scales is observed, while complete decoupling of the observables from the parameters occurs at high scales), 
non-Gaussian uncertainties as well as one-sided constraints, that in addition might constrain the model only 
very weakly. Also, counting the number of relevant observables $n$ might be non-trivial: For instance, after the 
discovery of the Higgs boson at the LHC, the limits on different Higgs masses set by 
the LEP experiments are expected to contribute only very weakly (if at all) to the total $\chi^{2}$ in a fit of the 
cMSSM. This is because the measurements at the LHC indicate that the lightest Higgs Boson has a mass 
significantly higher than the lower mass limit set by LEP. In such a situation, it is not 
clear how much such a one-sided limit actually is expected to contribute to the distribution of $\chi^2$ values.

For the above reasons, the accurate determination of the $p$-values for the fits presented in this paper requires 
the consideration of pseudo experiments or ``toy observable sets''.  Under the assumption that a particular 
best fit point provides an accurate description of nature, pseudo measurements are generated for each observable.  
Each pseudo measurement is based on the best fit prediction for the respective observable, taking into account 
both the absolute uncertainty on the best fit point, as well as the shape of the underlying probability density function. 
For one unique set of pseudo measurements, the fit is repeated, and a new best fit point is determined with a new 
minimum $\chi^{2}_{\mathrm{BF},i}$.

This procedure is repeated $n_{\textrm{toy}}$ times, and the number $n_{p}$ of fits using pseudo measurements with $\chi^{2}_{\mathrm{BF},i} \geq \chi^{2}_{\mathrm{BF}}$ is 
determined. The $p$-value is then given by the fraction 
\begin{eqnarray}
p = \frac{n_{p}}{n_{\textrm{toy}}} .
\end{eqnarray}
This procedure requires a considerable amount of CPU time; the number of sets of pseudo measurements is thus limited and the resulting $p$-value is subject to a statistical uncertainty. Given the true $p$-value,
\begin{eqnarray}
p_{\infty} = \lim_{n_{\textrm{toy}} \rightarrow \infty} p,
\end{eqnarray}
$n_{p}$ varies according to a binomial distribution $B(n_{p}|p_{\infty}, n_{\textrm{toy}})$, which in a rough approximation gives an uncertainty of 
\begin{eqnarray}
\Delta p = \sqrt{\frac{p\cdot (1-p)}{n_{\textrm{toy}}}}
\end{eqnarray}
on the $p$-value.

\subsubsection{Generation of Pseudo Measurements for the cMSSM}
In the present fit of the cMSSM a few different classes of observables have been used and the pseudo experiments have been generated accordingly.
In this work we distinguish different smearing procedures for the observables:
\begin{enumerate}
\item[a)] For a Gaussian observable with best fit prediction $O_{i}^{BF}$ and an 
absolute uncertainty $\sigma_{i}^{BF}$ at the best fit point, pseudo measurements have been generated by throwing a random number according to the probability density function
\begin{eqnarray}
P\left( O_{i}^{\textrm{toy}}\right) = \frac{1}{\sqrt{2\pi} \sigma_{i}^{BF}} \cdot \exp\left( - \frac{\left(O_{i}^{\textrm{toy}} - O_{i}^{BF} \right)^2 }{2 {\sigma_{i}^{BF}}^2}\right ).
\end{eqnarray}

\item[b)] For the measurements of the Higgs signal strengths and the Higgs mass, the smearing has been performed by means of the covariance matrix at the best fit point.

\item[c)] For the ATLAS 0-Lepton search~\cite{Aad:2014wea} (see Section~\ref{sec:observables:measurements}), the 
number of observed events has been smeared according to a Poisson distribution.  The expectation value of 
the Poisson distribution has been generated for each toy by taking into account the nominal value and the 
systematic uncertainty on both the background and signal expectation at the best fit point. The systematic 
uncertainties are assumed to be Gaussian.

\item[d)] The best fit point for each set of observables features a lightest Higgs boson with a mass well above the LEP 
limit. Assuming the best fit point, the number of expected Higgs events in the LEP searches is therefore 
negligible and has been ignored. For this reason, the LEP limit has been smeared directly assuming a Gaussian 
probability density function.
\end{enumerate}
\subsubsection{Rerunning the Fit}
Due to the enormous amount of CPU time needed to accurately sample the parameter space of the cMSSM 
and calculate a set of predictions at each point, a complete resampling for each set of pseudo measurements 
is prohibitive. 

For this reason the pseudo fits have been performed using only the points included in the original Markov chain, for which all necessary predictions have been calculated in the original scan.

In addition, an upper cut on the $\chi^{2}$ (calculated with respect to the real measurements) of $\Delta\chi^{2}
 \leq 15$ has been applied to further reduce CPU time consumption. The cut is motivated by the fact, that in order to find a toy best fit point that far from the original minimum, the outcome of the pseudo measurements would have to be extremely unlikely.
While this may potentially prevent a pseudo fit from finding the true minimum, tests with completely 
Gaussian toy models have shown that the resulting $\chi^{2}$ distributions perfectly match the expected 
$\chi^{2}$ distribution for all tested numbers of degrees of freedom.

As will be shown in Section~\ref{sec:results:toy}, in general we observe a trend towards less pseudo data fits with high $\chi^2$ values in the upper tail of the distribution than expected from the naive gaussian case. This further justifies that the  $\Delta\chi^{2} \leq 15$ cannot be expected to bias the full result of the pseudo data fits.

Nevertheless, the $p$-value calculated using the described procedure may be regarded as conservative in the sense that the true $p$-value may very well be even lower. 
Hence, if it is found below a certain threshold of e.g.{} 5\%, it is not expected that there is a bias that the true $p$-value 
for infinite statistics is found at larger values. If for a particular toy fit the true best fit point is not 
included in the original Markov chain, the minimum $\chi^2$ for that pseudo fit will be larger than the true minimum for 
that pseudo fit, which artificially increases the $p$-value.

%% file: observables.tex
\section{Observables}\label{sec:observables}
The parameters of the cMSSM are constrained by  precision observables, like 
$(g-2)_\mu$, astrophysical observations including in particular direct dark matter 
detection limits and the dark matter relic density, by collider sear\-ches for 
supersymmetric particles and by the properties of the Higgs boson. In this section we 
describe the observables that enter our fits. The measurements are given in Section~\ref{sec:observables:measurements} while the codes used to obtain the corresponding model 
predictions are described in Section~\ref{sec:observables:predictions}. 

\input{observables_measurements}
\input{observables_predictions}

%% file: observables_measurements.tex
\subsection{Measurements and exclusion limits}\label{sec:observables:measurements}
We employ the same set of precision observables as in our previous
analysis Ref.~\cite{Bechtle:2012zk}, but with updated measurements as
listed in Tab.\,\ref{tab:measurements}.  They include the anomalous
magnetic moment of the muon $(g-2)_\mu \equiv a_\mu$, the effective
weak mixing angle $\sin^2 \theta_\mathrm{eff}$, the masses of the top
quark and $W$ boson, the $B_s$ oscillation frequency $\Delta m_s$, as
well as the branching ratios $\bsmm$, $\btaunu$, and $\bsgamma$. The
Standard Model parameters that have been fixed are collected in
Tab.\,\ref{tab:smparameters}. Note that the top quark mass $m_t$ is
used both as an observable, as well as a floating parameter in the
fit, since it has a åsignificant correlation especially with the light
Higgs boson mass.

\begin{table}[t]
\caption{ Precision observables used in the 
  fit.}\label{tab:measurements}
\begin{tabular}{llc}
\hline\noalign{\smallskip}
      $ a_{\mu}-a_{\mu}^{\mathrm{\mathrm{SM}}}$ & $(28.7 \pm 8.0 )\times10^{-10}$      & \cite{Bennett:2006fi,Davier:2010nc}\\
      $ \sin^2\theta_{\mathrm{eff}}$            & $0.23113 \pm 0.00021 $          & \cite{ALEPH:2005ab}\\
      $m_t$                       & $(173.34 \pm 0.27 \pm 0.71)$\,GeV         & \cite{ATLAS:2014wva}\\                                                     
      $m_W$                                     & $(80.385 \pm 0.015 )$\,GeV       & \cite{Group:2012gb} \\
       $ \Delta m_{s}$                           & $(17.719 \pm  0.036 \pm 0.023 )\,\mathrm{ps}^{-1}$  & \cite{Beringer:1900zz} \\
      $ {\cal B}(B_s\to\mu\mu) $                & $(2.90 \pm 0.70 )\times 10^{-9}$                   & \cite{CMSandLHCbCollaborations:2013pla}\\
      $ {\cal B}(b\to s\gamma)$ & $(3.43 \pm  0.21 \pm 0.07 )\times 10^{-4}$       & \cite{Amhis:2012bh}\\
      $ {\cal B}(B\to\tau\nu)$   & $(1.05 \pm 0.25 ) \times 10^{-4}$                  & \cite{Beringer:1900zz}\\
\noalign{\smallskip}\hline 
\end{tabular}
\end{table}

\begin{table}[t]
\caption{Standard Model parameters that have been fixed. Please note
  that $m_b$ and $m_c$ are $\overline{\textrm{MS}}$ masses at their respective mass scale,
  while for all other particles on-shell masses are used.}
\label{tab:smparameters}
\begin{tabular}{lll}
\hline\noalign{\smallskip}
      $ 1/\alpha_{\rm em}  $ & $128.952$     &  \cite{Davier:2010nc} \\
      $ G_{\rm F} $ & $(1.1663787\times 10^{-5})$~GeV$^{-2}$  &  \cite{Beringer:1900zz}  \\
      $ \alpha_{\rm s}$ &   $0.1184$   &  \cite{Beringer:1900zz}  \\
      $ m_Z$ &  $91.1876$~GeV &  \cite{Beringer:1900zz}  \\
      $ m_b$ &  $4.18$~GeV  &  \cite{Beringer:1900zz}  \\
      $ m_{\tau}        $ &  $1.77682$~GeV &  \cite{Beringer:1900zz}  \\
      $ m_c  $           &   $1.275$ ~GeV &  \cite{Beringer:1900zz}  \\
\noalign{\smallskip}\hline 
\end{tabular}
\end{table}

Dark matter is provided by the lightest supersymmetric particle, which
we require to be solely made up of the neutralino. We use the dark
matter relic density $\Omega h^2=0.1187 \pm 0.0017$ as obtained by the
Planck collaboration~\cite{Ade:2013zuv} and bounds on the
spin-independent WIMP--nucleon scattering cross section as measured by
the LUX experiment~\cite{Akerib:2013tjd}.

Supersymmetric particles have been searched for at the LHC in a
plethora of final states. In the cMSSM parameter region preferred by
the precision observables listed in Tab.~\ref{tab:measurements}, the
LHC jets plus missing transverse momentum searches provide the
strongest constraints.  We thus implement the ATLAS analysis of
Ref.\,\cite{Aad:2014wea} in our fit, as described in some detail in~\cite{Bechtle:2012zk}. Furthermore we enforce the LEP bound on the
chargino mass, $m_{\tilde{\chi}^\pm_1}>103.5$\,GeV\,\cite{LEPSUSYWG}. The constraints from Higgs searches at LEP are
incorporated through the $\chi^2$ extension provided by
\textsc{HiggsBounds
  4.1.1}~\cite{Bechtle:2008jh,Bechtle:2011sb,Bechtle:2013gu,Bechtle:2013wla},
which also provides limits on additional heavier Higgs bosons. The
signal rate and mass measurements of the experimentally established
Higgs boson at $125$~GeV are included using the program
\textsc{HiggsSignals 1.2.0}~\cite{Bechtle:2013xfa} (see also
Ref.~\cite{Bechtle:2014ewa} and references
therein). \textsc{HiggsSignals} is a general tool which allows the
test of any model with Higgs-like particles against the measurements
at the LHC and the Tevatron.  Therefore, its default observable set of
Higgs rate measurements is very extensive. As is discussed in detail
in Section~\ref{sec:results:toy}, this provides maximal flexibility
and sensitivity on the constraints of the allowed parameter ranges,
but is not necessarily ideally tailored for goodness-of-fit
tests. There, it is important to combine observables which the model
on theoretical grounds cannot vary independently. In order to take
this effect into account, in our analysis we compare five different
Higgs observable sets:
\begin{list}{\it Set \arabic{setcounter}\!}{\usecounter{setcounter}}\setlength\itemsep{1em}
\item \textit{(Large Observable Set):} This set is the default
  observable set provided with \textsc{HiggsSignals 1.2.0}, containing
  in total 80 signal rate measurements obtained from the LHC and
  Tevatron experiments. It contains all available sub-channel /
  category measurements in the various Higgs decay modes investigated
  by the experiments. Hence, while this set is most appropriate for
  resolving potential deviations in the Higgs boson coupling
  structure, it comes with a high level of redundancy. Detailed
  information on the signal rate observables in this set can be found
  in Ref.~\cite{Bechtle:2014ewa}. Furthermore, the set contains 4 mass
  measurements, performed by ATLAS and CMS in the $h\to \gamma\gamma$
  and $h\to ZZ^{(*)}\to4\ell$ channels. It is used as a cross-check
  for the derived observable sets described below.
\item \textit{(Medium
    Observable Set):} This set contains 10 inclusive rate
  measurements, performed in the channels $h\to \gamma\gamma$,
  $h\to ZZ$, $h\to WW$, $Vh\to V(bb)$ ($V=W,Z$), and $h\to \tau\tau$
  by ATLAS and CMS, listed in Tab.~\ref{tab:MedObsSet}. As in
  \textit{Set~1}, 4 Higgs mass measurements are included.  
\item
  \textit{(Small Observable Set):} In this set, the
  $h\to \gamma\gamma$, $h\to~ZZ$ and $h\to WW$ channels are combined
  to a measurement of a universal signal rate, denoted
  $h\to \gamma\gamma, ZZ, WW$ in the following. Together with the
  $Vh\to~V(bb)$ and $h\to \tau\tau$ from \textit{Set 2}, we have in
  total 6 rate measurements. Furthermore, in each LHC experiment the
  Higgs mass measurements are combined. The observables are listed in
  Tab.~\ref{tab:SmallObsSet}.  
\item \textit{(Combined Observable Set):} In this set we further
  reduce the number of Higgs observables by combining the ATLAS and
  CMS measurements for the Higgs decays to electroweak vector bosons
  ($V=W,Z$), photons, $b$-quarks and $\tau$-leptons. These
  combinations are performed by fitting a universal rate scale factor
  $\mu$ to the relevant observables from within
  \textit{Set~1}. Furthermore, we perform a combined fit to the Higgs
  mass observables of \textit{Set~1}, yielding
  $m_h = (125.73\pm 0.45)$\,GeV.\footnote{Note that the
    computing time needed for creating the pseudo-data fits presented
    in Section~\ref{sec:results:toy} means that the fits were starting
    to be performed significantly before the combined measurement of
    the Higgs boson mass $m_{h_{comb}}=125.09\pm0.21$\,GeV by the
    ATLAS and CMS collaborations was published~\cite{Aad:2015zhl}. We
    therefore performed our own combination, based on earlier results
    as published
    in~\cite{Aad:2013wqa,Chatrchyan:2013mxa,CMS:yva}. Given the
    applied theory uncertainty on the Higgs mass prediction of
    $\Delta m_{h_{theo}}=3$\,GeV, a shift of $0.64$\,GeV in the Higgs
    boson mass has a very small effect of
    $\Delta\chi^2\approx{\cal O}(0.64^2/3^2)=0.046$, which is
    negligible in terms of the overall conclusions in this paper.
  } 
  The observables of this set are listed in Tab.~\ref{tab:CombObsSet}.
\item \textit{(Higgs mass
    only):} Here, we do not use any Higgs signal rate measurements. We
  only use one combined Higgs mass observable, which in our case is
  $m_h = (125.73\pm 0.45)$\,GeV (see above).
\end{list}

\begin{table}[t]
\caption{Higgs boson mass and rate observables of \textit{Set 2} (Medium Observable Set).} 
\label{tab:MedObsSet}
\renewcommand{\arraystretch}{1.2}
\begin{tabular}{lcc}
\toprule
Experiment, Channel  &  observed $\mu$  &  observed $m_h$ \\
\midrule
ATLAS, $h\to WW\to \ell\nu\ell\nu$~\cite{Aad:2013wqa} & $0.99\substack{+0.31 \\ -0.28}$ & - \\
ATLAS, $h\to ZZ\to 4\ell$~\cite{Aad:2013wqa} & $1.43\substack{+0.40 \\ -0.35}$ & $(124.3\pm 1.1)$\,GeV  \\
ATLAS, $h\to \gamma\gamma$~\cite{Aad:2013wqa} & $1.55\substack{+0.33 \\ -0.28}$ & $(126.8\pm 0.9)$\,GeV \\
ATLAS, $h\to \tau\tau$~\cite{ATLAS:2012dsy} & $1.44\substack{+0.51 \\ -0.43}$ & - \\
ATLAS, $Vh\to V(bb)$~\cite{TheATLAScollaboration:2013lia} & $0.17\substack{+0.67 \\ -0.63}$ & - \\
\midrule
CMS, $h\to WW\to \ell\nu\ell\nu$~\cite{Chatrchyan:2013iaa} & $0.72\substack{+0.20 \\ -0.18}$ & -\\
CMS, $h\to ZZ\to 4\ell$~\cite{Chatrchyan:2013mxa} & $0.93\substack{+0.29 \\ -0.25}$ & $(125.6\pm 0.6)$\,GeV \\
CMS, $h\to \gamma\gamma$~\cite{CMS:yva} & $0.77\substack{+0.30 \\ -0.27}$ & $(125.4\pm 1.1)$\,GeV\\
CMS, $h\to \tau\tau$~\cite{Chatrchyan:2014vua} & $0.78\substack{+0.27 \\ -0.27}$ & - \\
CMS, $Vh\to V(bb)$~\cite{Chatrchyan:2014vua} & $1.00\substack{+0.50 \\ -0.50}$ & - \\
\bottomrule
\end{tabular}
\end{table}

\begin{table}[t]
\caption{Higgs boson mass and rate observables of \textit{Set 3} (Small Observable Set).}\label{tab:SmallObsSet}
\begin{threeparttable}
\renewcommand{\arraystretch}{1.2}
\begin{tabular}{lcc}
\toprule
Experiment, Channel  &  observed $\mu$  &  observed $m_h$  \\
\midrule
ATLAS, $h\to WW,ZZ,\gamma\gamma$~\cite{Aad:2013wqa} & $1.33\substack{+0.21 \\ -0.18}$ & $(125.5\pm 0.8)$\,GeV \\
ATLAS, $h\to \tau\tau$~\cite{ATLAS:2012dsy} & $1.44\substack{+0.51 \\ -0.43}$ & -  \\
ATLAS, $Vh\to V(bb)$~\cite{TheATLAScollaboration:2013lia} & $0.17\substack{+0.67 \\ -0.63}$ & -  \\
\midrule
CMS, $h\to WW,ZZ,\gamma\gamma$ \tnote{$\dagger$} & $0.80\substack{+0.16 \\ -0.15}$ & $(125.7\pm 0.6)$\,GeV \\
CMS, $h\to \tau\tau$~\cite{Chatrchyan:2014vua} & $0.78\substack{+0.27 \\ -0.27}$ & -  \\
CMS, $Vh\to V(bb)$~\cite{Chatrchyan:2014vua} & $1.00\substack{+0.50 \\ -0.50}$ & -  \\
\bottomrule
\end{tabular}
\begin{tablenotes}
\footnotesize
\item[$\dagger$]The combination of the CMS $h\to WW,ZZ,\gamma\gamma$ channels has been performed with \textsc{HiggsSignals} using results from Ref.~\cite{CMS:ril,Chatrchyan:2013iaa,Chatrchyan:2013mxa}. The combined mass measurement for CMS is taken from Ref.~\cite{CMS:yva}.
\end{tablenotes}
\end{threeparttable}
\end{table}

\begin{table}[t]
\caption{Higgs boson mass and rate observables of \textit{Set 4} (Combined Observable Set).}
\label{tab:CombObsSet}
\renewcommand{\arraystretch}{1.2}
\begin{tabular}{lcc}
\toprule
Experiment, Channel  &  observed $\mu$  &  observed $m_h$ \\
\midrule
ATLAS+CMS, $h\to WW,ZZ$ & $0.94\substack{+0.17 \\ -0.16}$ & \multirow{2}{*}{$(125.73\pm 0.45)$\,GeV} \\
ATLAS+CMS, $h\to \gamma\gamma$ & $1.16\substack{+0.22 \\ -0.20}$ &  \\
ATLAS+CMS, $h\to \tau\tau$ & $1.11\substack{+0.24 \\ -0.23}$ & - \\
ATLAS+CMS, $Vh,tth\to bb$ & $0.69\substack{+0.37 \\ -0.37}$ & - \\
\bottomrule
\end{tabular}
\end{table}

%% file: observables_predictions.tex
\subsection{Model predictions}\label{sec:observables:predictions}

We use the following public codes to calculate the predictions for the
relevant observables: The spectrum is calculated with \textsc{SPheno
  3.2.4}~\cite{Porod:2003um,Porod:2011nf}. First the two-loop RGEs~\cite{Martin:1993zk} are used to obtain the parameters at the scale
$Q=\sqrt{m_{\tilde t_1} m_{\tilde t_2}}$.  At this scale the complete
one-loop corrections to all masses of supersymmetric particles are
calculated to obtain the on-shell masses from the underlying
$\overline{\rm DR}$ parameters~\cite{Pierce:1996zz}. A measure of the
theory uncertainty due to the missing higher-order corrections is
given by varying the scale $Q$ between $Q/2$ and $ 2 Q$. We find that
the uncertainty on the strongly interacting particles is about 1-2\%,
whereas for the electroweakly interacting particles it is of order a
few per mille~\cite{Porod:2011nf}.

Properties of the Higgs bosons as well as $a_{\mu}$, $\Delta m_s$,
$\sin^2\theta_\mathrm{eff}$ and $m_W$ are obtained with
\textsc{FeynHiggs 2.10.1}~\cite{Heinemeyer:1998yj}, which -- compared
to \textsc{FeynHiggs 2.9.5} and earlier versions -- contains a
significantly improved calculation of the Higgs boson mass~\cite{Hahn:2013ria} for the case of a heavy SUSY spectrum. This
improves the theoretical uncertainty on the Higgs mass calculation
from about 3-4 GeV in cMSSM scenarios~\cite{Degrassi:2002fi,Allanach:2004rh,Heinemeyer:2004xw} to about 2\,GeV~\cite{Buchmueller:2013psa}.

The B-physics branching ratios are calculated by \textsc{SuperIso 3.3}~\cite{Mahmoudi:2008tp}. We have checked that the results for the
observables, that are also implemented in \textsc{SPheno} agree within
the theoretical uncertainties, see also~\cite{Porod:2014xia} for a
comparison with other codes.

For the astrophysical observables we use \textsc{MicrOMEGAs 3.6.9}~\cite{Belanger:2001fz, Belanger:2004yn} to calculate the dark matter
relic density and \textsc{DarkSUSY 5.0.5}~\cite{Gondolo:2004sc} via
the interface program \textsc{AstroFit}~\cite{AstroFit} for the direct
detection cross section.

\begin{figure}
  \includegraphics[width=0.5\textwidth]{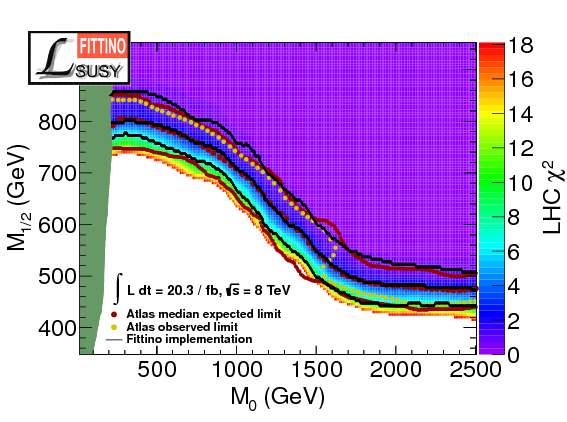}
  \caption{Comparison of the emulation of the ATLAS 0-Lepton search with the published ATLAS result. In red dots 
 we show the Atlas median expected limit; the red lines denote the corresponding 1$\sigma$ uncertainty. The central black line 
 is the result of the \textsc{Fittino} implementation described in the text. The upper and lower black curves are the corresponding 
 1$\sigma$ uncertainty. The yellow dots are the observed Atlas  limit.}
  \label{fig:atlas0lepton} 
\end{figure}

For the calculation of the expected number of signal events in the
ATLAS jets plus missing transverse momentum search, we use the Monte
Carlo event generator \textsc{Herwig++}~\cite{Bahr:2008pv} and a
carefully tuned and validated version of the fast parametric detector
simulation \textsc{Delphes}~\cite{Ovyn:2009tx}. For $\tan \beta = 10$
and $A_{0} = 0$, a fine grid has been produced in $M_{0}$ and
$M_{1/2}$. In addition, several smaller, coarse grids have been
defined in $A_{0}$ and $\tan \beta$ for fixed values of $M_{0}$ and
$M_{1/2}$ along the expected ATLAS exclusion curve to correct the
signal expectation for arbitrary values of $A_{0}$ and $\tan\beta$. We
assume a systematic uncertainty of $10\%$ on the expected number of
signal events. In Fig.~\ref{fig:atlas0lepton} we compare the
expected and observed limit as published by the ATLAS collaboration to
the result of our emulation. The figure shows that the procedure works
reasonably well and is able to reproduce with sufficient precision the
expected exclusion curve, including the $\pm 1\sigma$ variations.

We reweight the events depending on their production channel according
to NLO cross sections obtained from \textsc{Prospino}~\cite{Beenakker:1996ed,Beenakker:1996ch,Beenakker:1997ut}. Renormalisation and factorisation scales have
been chosen such that the NLO+NLL resummed cross section
normalisations~\cite{Kulesza:2008jb, Beenakker:2009ha,
  Beenakker:2010nq, Beenakker:2011fu, Kramer:2012bx} are reproduced for squark
and gluino production.

For all predictions we take theoretical uncertainties into account,
most of which are parameter dependent and reevaluated at every point
in the MCMC. For the precision observables, they are given in Tab.~\ref{tab:theounc}. For the dark matter relic density we assume a
theoretical uncertainty of $10\%$, for the neutralino-nucleon cross
section entering the direct detection limits we assign $50\%$
uncertainty (see Ref.~\cite{Bechtle:2012zk} for a discussion of this
uncertainty arinsing from pion-nuclueon form factors), for the Higgs
boson mass prediction $2.4\%$, and for Higgs rates we use the
uncertainties given by the LHC Higgs Cross Section Working Group~\cite{Heinemeyer:2013tqa}.

\begin{table}
\caption{Theoretical uncertainties of the precision observables used in the fit.}\label{tab:theounc}
\begin{tabular}{lc}
\hline\noalign{\smallskip}
      $ a_{\mu}-a_{\mu}^{\mathrm{\mathrm{SM}}}$   & $7\%$    \\
      $ \sin^2\theta_{\mathrm{eff}}$                 & $0.05\%$    \\
      $m_t$                       & 1\,GeV  \\                                                       
      $m_W$                                     & $0.01\%$  \\
      $ \Delta m_{s}$                          & $24\%$   \\
            $ {\cal B}(B_s\to\mu\mu) $                & $26\%$                     \\
      $ {\cal B}(b\to s\gamma)$  & $14\%$   \\
      $ {\cal B}(B\to\tau\nu)$      & $20\%$        \\    
\noalign{\smallskip}\hline 
\end{tabular}
\end{table}

One common challenge for computing codes specifically developed for
SUSY predictions is that they might not always exactly predict the
most precise predictions of the SM value in the decoupling limit. The
reason is that specific loop corrections or renormalisation
conventions are not always numerically implemented in the same way, or
that SM loop contributions might be missing from the SUSY
calculation. In most cases these differences are within the theory
uncertainty, or can be used to estimate those. One such case of
interest for this fit occurs in the program \textsc{FeynHiggs}, which does not
exactly reproduce the SM Higgs decoupling
limit~\cite{PrivateCommunicationSven} as used by the LHC Higgs
cross-section working group~\cite{Heinemeyer:2013tqa}. To compensate
this, we rescale the Higgs production cross sections and partial
widths of the SM-like Higgs boson. We determine the scaling factors by
the following procedure~\cite{PrivateCommunicationSven}: We fix
$\tan\beta =10$. We set all mass parameters of the MSSM (including the
parameters $\mu$ and $m_A$ of the Higgs sector) to a common value
$m_\mathrm{SUSY}$. We require all sfermion mixing parameters $A_f$ to
be equal. We vary them by varying the ratio $X_t/m_ \mathrm{SUSY}$,
where $X_t = A_t - \mu/\tan\beta$.  The mass of the Higgs boson
becomes maximal for values of this ratio of about $\pm 2 $.  We scan
the ratio between these values. In this way we find for each
$m_\mathrm{SUSY}$ two parameter points which give a Higgs boson mass
of about 125.5\,GeV.  One of these has negative $X_t$, the other
positive $X_t$. We then determine the scaling factor by requiring that
for $m_\mathrm{SUSY}=4$\,TeV and negative $X_t$ the production cross
sections and partial widths of the SM-like Higgs boson are the same as
for a SM Higgs boson with the same mass of 125.5\,GeV. We then
determine the uncertainty on this scale factor by comparing the result
with scale factors which we would have gotten by choosing
$m_\mathrm{SUSY}=3$\,TeV, $m_ \mathrm{SUSY}=5$\,TeV or a positive sign
of $X_t$. This additional uncertainty is taking into account in the
$\chi^2$ computation. By this procedure we derive scale factors
between $0.95$ and $1.23$ with uncertainties of less than $0.6\%$.

%% file: results.tex
\section{Results}\label{sec:results}

In Section~\ref{sec:results:pl}, we show results based on the simplistic and common profile
likelihood technique, which all frequentist fits, including us, have hitherto been employing. 
In Section~\ref{sec:results:pl:vacuum} a full scan of the allowed parameter space for a 
stable vacuum is shown, before moving on to novel results from toy fits in Section~\ref{sec:results:toy}.

\input{results_pl}

\input{results_pl_vacuum}
\input{results_toy}

%% file: results_pl.tex
\subsection{Profile likelihood based results}\label{sec:results:pl}

\begin{figure*}[t]
  \begin{center}
    \subfigure[1$\,\sigma$ and 2$\,\sigma$ contour plot in in the ($M_{0}$,$\,M_{1/2}$)--plane for the Small Observable Set.]{
      \centering\includegraphics[width=0.45\textwidth]{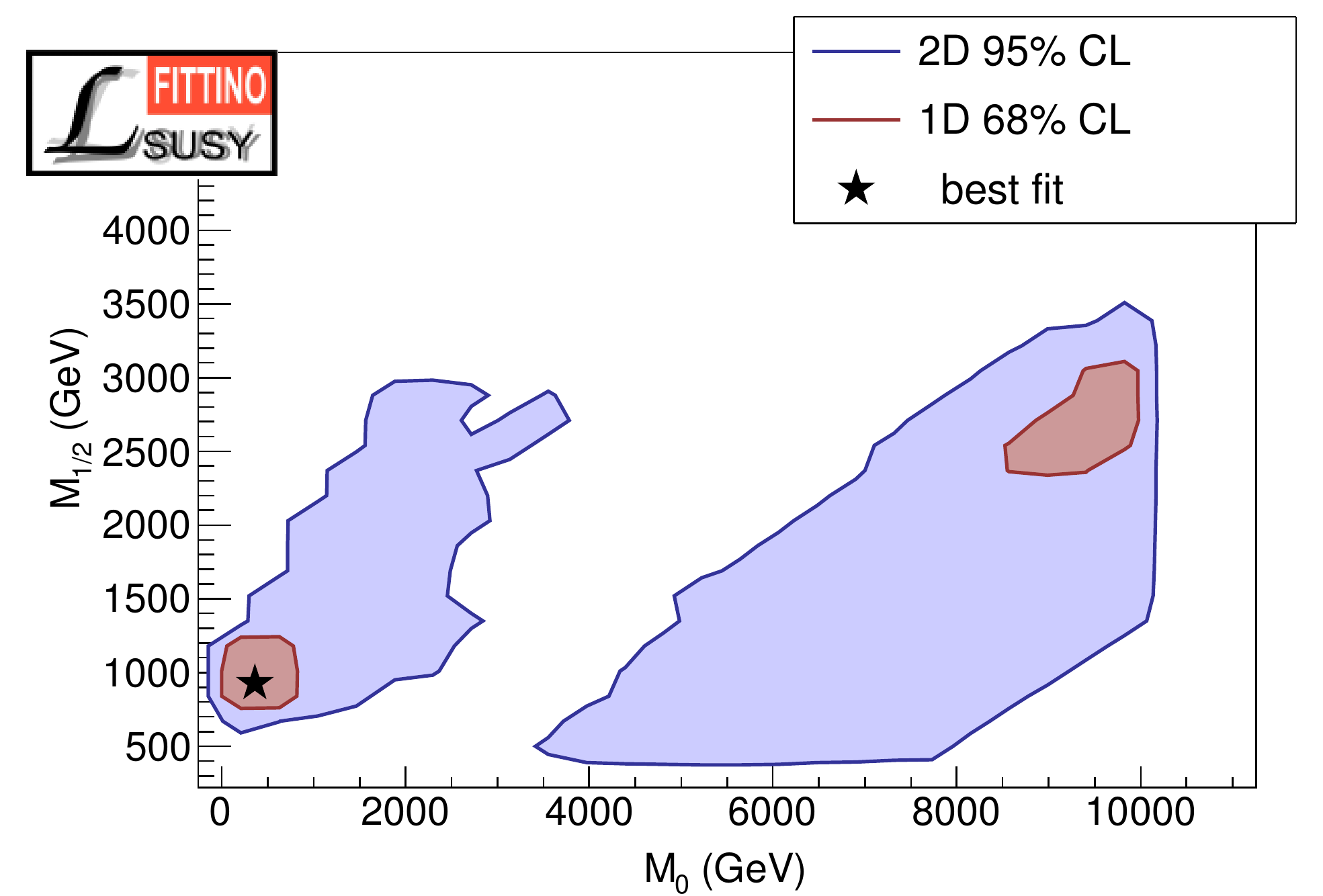}
      \label{fig:SmallObsSet:PL:M0M12}  
    }
    \subfigure[1$\,\sigma$ and 2$\,\sigma$ contour plot in in the ($M_{0}$,$\,M_{1/2}$)--plane 
        for the Medium Observable Set.]{
      \centering\includegraphics[width=0.45\textwidth]{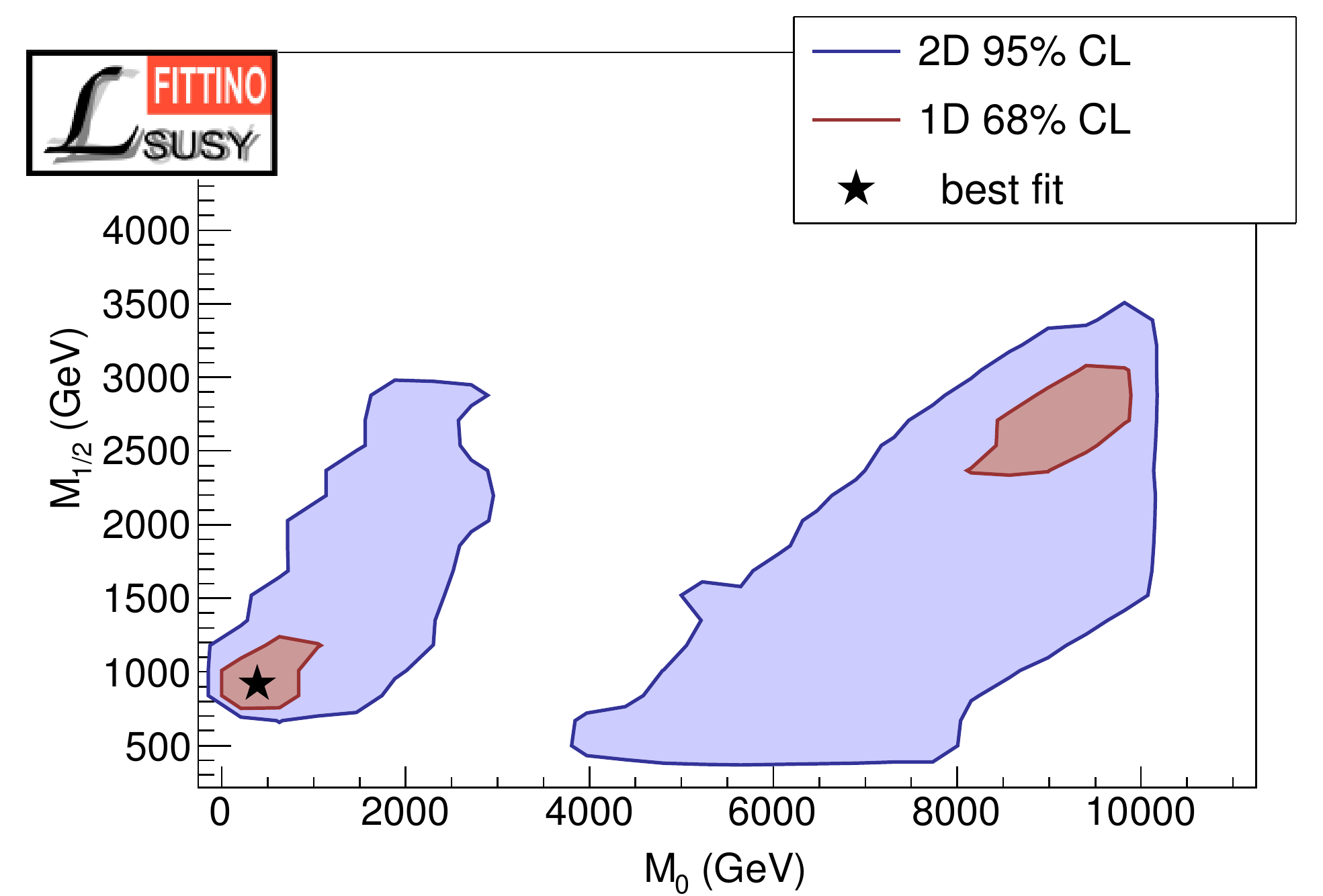}
      \label{fig:MediumObsSet:PL:M0M12}  
    }
    \subfigure[1$\,\sigma$ and 2$\,\sigma$ contour plot in the ($M_{0}$,$\,M_{1/2}$)--plane for the \mbox{Large Observable Set}.]{
      \centering\includegraphics[width=0.45\textwidth]{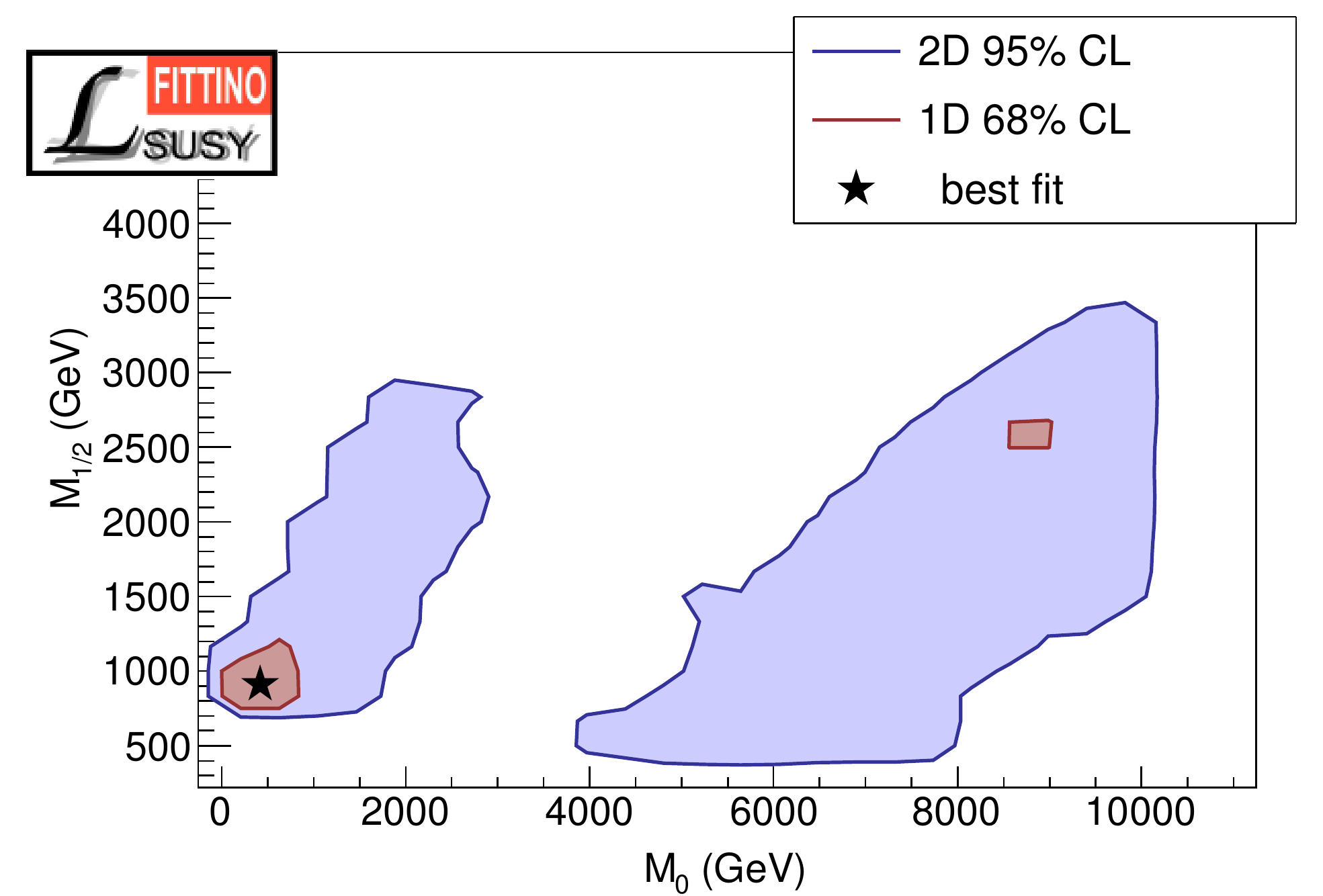}
      \label{fig:LargeObsSet:PL:M0M12}  
    }
    \subfigure[1$\,\sigma$ and 2$\,\sigma$ contour plot in the ($A_{0},\,\tan\beta$)--plane for the Medium Observable Set.]{
      \centering\includegraphics[width=0.45\textwidth]{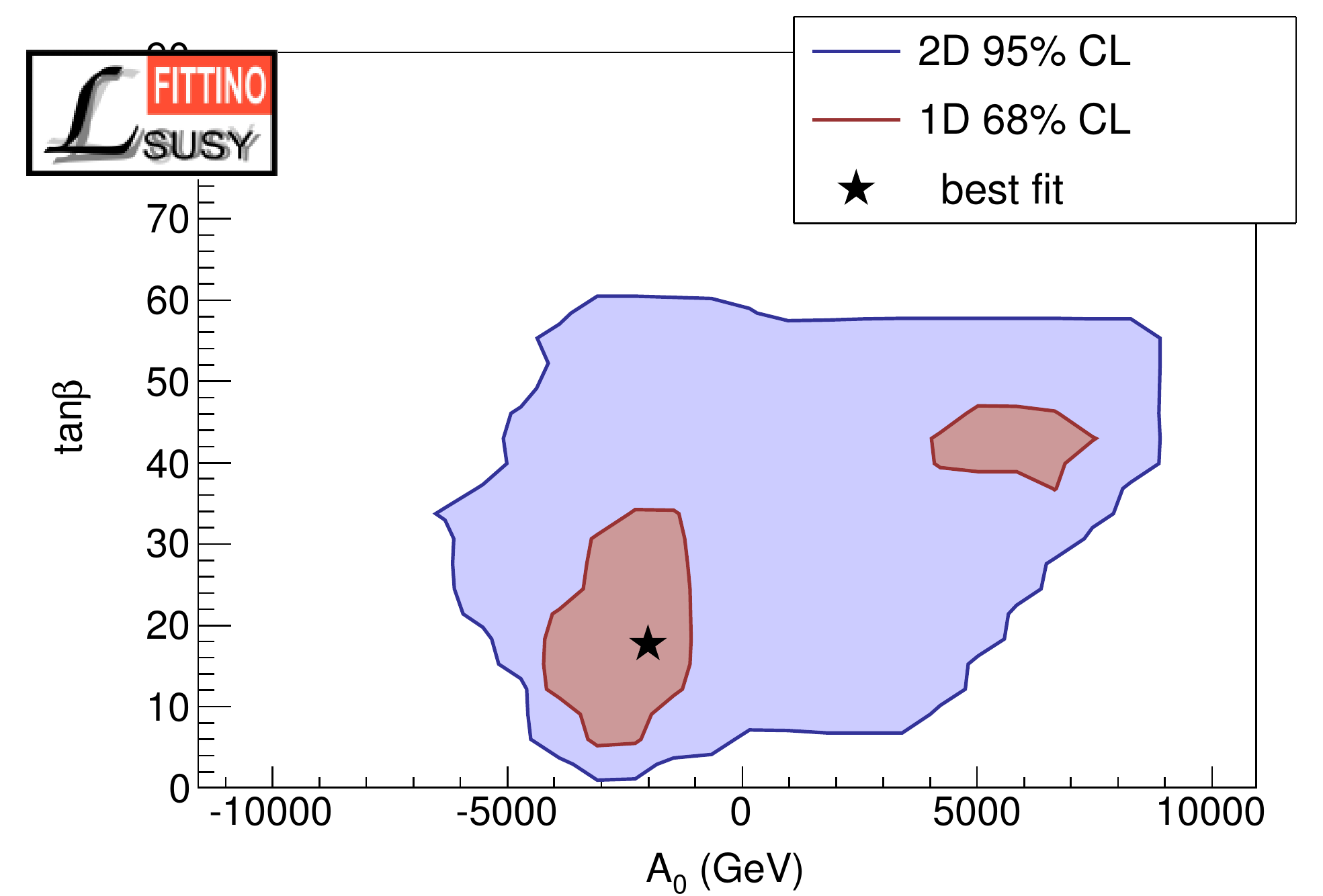}
      \label{fig:MediumObsSet:PL:A0TanBeta}  
    }
  \end{center}
  \caption{1$\,\sigma$ and 2$\,\sigma$ contour plots for different
    projections and different observable sets. It can be seen that the preferred parameter
    region does not depend on the specific observable set.}\label{fig:PL:projections}
\end{figure*}

In this section we describe the preferred parameter space region of
the cMSSM and its physical properties.  Since a truly complete
frequentist determination of a confidence region would require not
only to perform toy fits around the best fit point (as described in
Section~\ref{sec:methods:pvalue} and \ref{sec:results:toy}) but around
\textit{every} cMSSM parameter point in the scan, we rely here on the
profile likelihood technique. This means, we show various projections
of the 1D-$1\sigma$/1D-$2\sigma$/2D-$2\sigma$ regions defined as
regions which satisfy $\Delta\chi^2<1/4/5.99$ respectively. In this
context, profile likelihood means that out of the 5 physical
parameters in the scan, the parameters not shown in a plot are
profiled, \textit{i.e.} a scan over the hidden dimensions is performed
and for each selected visible parameter point the lowest $\chi^2$
value in the hidden dimensions is chosen. Obviously, no systematics
nuisance parameters are involved, since all systematic uncertainties
are given by relative or absolute Gaussian uncertainties, as discussed
in Section~\ref{sec:methods}. One should keep in mind that this
correspondence is actually only exact when the observed distribution
of $\chi^2$ values in a set of toy fits is truly $\chi^2$-distributed,
which -- as discussed in Section~\ref{sec:results:toy} -- is not the
case.  Nevertheless, since the exact method is not computationally
feasible, this standard method, as used in the literature in all
previous frequentist results, gives a reasonable estimation of the
allowed parameter space. In Section~\ref{sec:results:toy} more
comparisons between the sets of toy fit results and the profile
likelihood result will be discussed.

Note that for the discussion in this and the next section,
we treat the region around the best fit point as ``allowed'', even
though, depending on the observable set, an exclusion of the complete
model will be discussed in Section~\ref{sec:results:toy}.

All five Higgs input parameterisations introduced in Section~\ref{sec:observables} lead to very 
similar results when interpreted with the profile likelihood technique.  As an example, Figures~\ref{fig:SmallObsSet:PL:M0M12} - \ref{fig:LargeObsSet:PL:M0M12} show the ($M_{0},\,
M_{1/2}$)--projection of the best fit point, 1D-$1\sigma$ and 2D-$2\sigma$ regions for the 
Small, the Large and the Medium Observable Set.
Thus, in the remainder of this section, we concentrate on results from the Medium Observable Set.

The ($M_0,\,M_{1/2}$)--projection in
Fig.~\ref{fig:MediumObsSet:PL:M0M12} shows two disjoint
regions. While in the region of the global $\chi^2$ minimum, values of
less than 900\,GeV for $M_0$ and less than 1300\,GeV for $M_{1/2}$ are
preferred at $1\sigma$, in the region of the secondary minimum values
of more than 7900\,GeV for $M_0$ and more than 2100\,GeV for $M_{1/2}$
are favored (see also Tab.~\ref{tab:pl:bf}).

\begin{table}[t]
\caption{Central values and $1\sigma$ uncertainties of the free model parameters at the global and secondary minimum when using the Medium Observable Set}
\label{tab:pl:bf}
\begin{tabular}{c|c|c}
Parameter & global minimum & secondary minimum \\
\hline
$M_0$ (GeV) & $387.4^{+481.7}_{-151.2}$ & $8983.4^{+990.6}_{-1039.6}$\\[6pt]
$M_{1/2}$ (GeV) & $918.2^{+297.7}_{-59.3}$ & $2701.1^{+582.6}_{-560.5}$\\[6pt]
$A_0$ (GeV) & $-2002.8^{+541.5}_{-1992.9}$ & $5319.0^{+2339.8}_{-1357.9}$\\[6pt]
$\tan\beta$ & $17.7^{+16.8}_{-10.8}$ & $43.2^{+5.5}_{-6.6}$\\[6pt]
$m_{t}$ (GeV) & $174.3^{+1.1}_{-1.1}$ & $172.1^{+0.6}_{-0.6}$\\[6pt]
\end{tabular}
\end{table}

\begin{figure}[t]
  \includegraphics[width=0.55\textwidth]{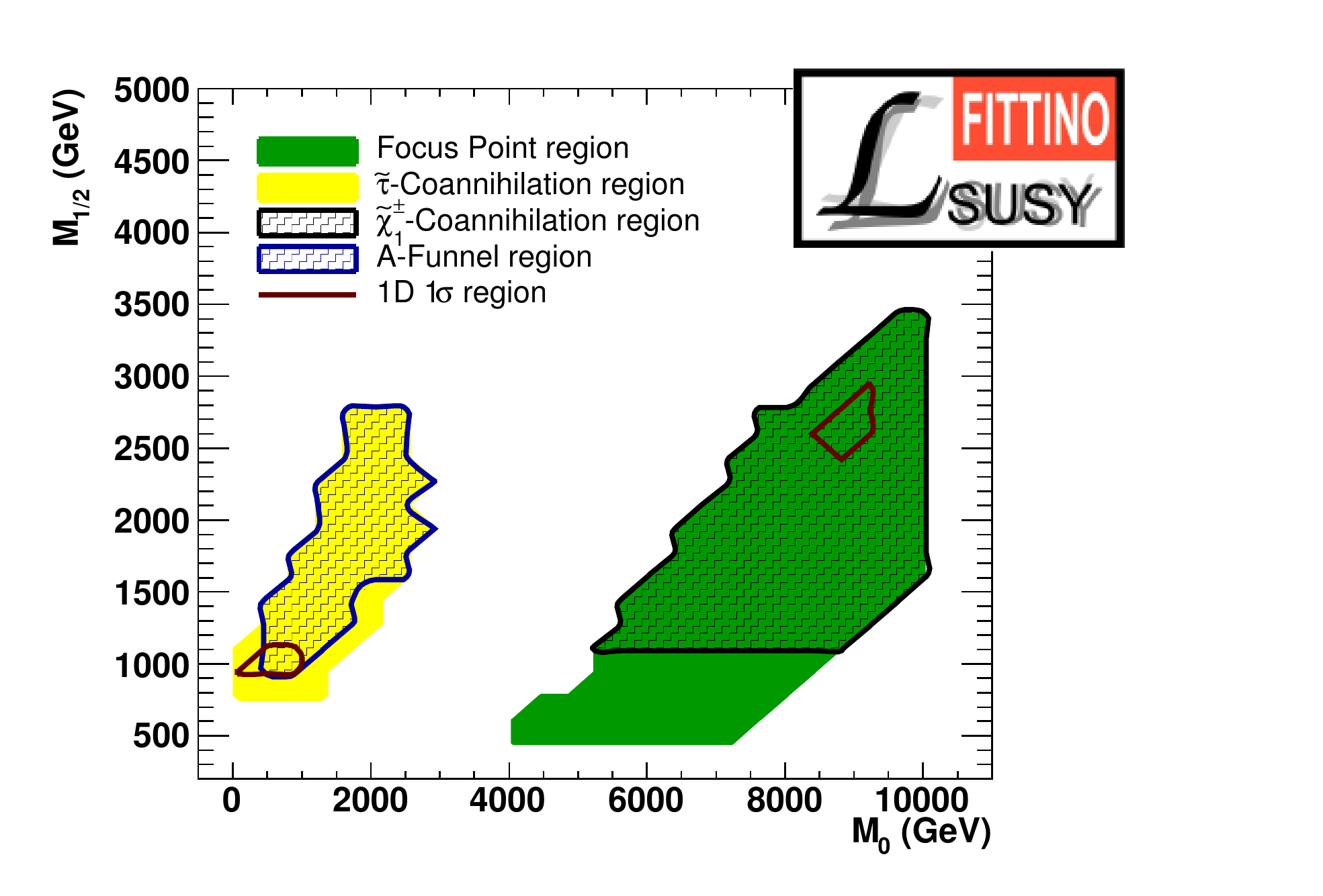}
  \caption{ 2$\,\sigma$ region in the ($M_{0},\,M_{1/2}$)--plane for the Medium Observable Set. Regions 
  with different dark matter annihilation mechanisms are indicated. The enclosed red areas denote the 
  best fit regions shown in Fig.~\ref{fig:MediumObsSet:PL:M0M12}.}
  \label{fig:MediumObsSet:PL:DarkMatterAnnihilation}  
\end{figure}

 The different regions are characterised by different dominant dark matter
annihilation mechanisms as shown in
Fig.~\ref{fig:MediumObsSet:PL:DarkMatterAnnihilation}. Here we
define the different regions similarly to
Ref.~\cite{Buchmueller:2014yva} by the following kinematical
conditions, which we have adapted such that each point of the
2D-$2\sigma$ region belongs to at least one region:

{ \setlength{\tabcolsep}{2pt}
\begin{tabular}{llll}
- & $\tilde{\tau}_1$ coannihilation: &${m_{\tilde{\tau}_1}}/{m_{\tilde{\chi}_1^0}}-1<0.15$\\
- & $\tilde{t}_1$ coannihilation: &${m_{\tilde{t}_1}}/{m_{\tilde{\chi}_1^0}}-1<0.2$\\
- & $\tilde{\chi}_1^\pm$ coannihilation:& ${m_{\tilde{\chi}_1^\pm}}/{m_{\tilde{\chi}_1^0}}-1<0.1$\\
- & $A/H$ funnel:& $|{m_{A}}/{2m_{\tilde{\chi}_1^0}}-1|<0.2$\\
- & focus point region: & $|{\mu}/{m_{\tilde{\chi}_1^0}}-1|<0.4$
\end{tabular}}

With these definitions each parameter point of the preferred
2D-$2\sigma$ region belongs either to the $\tilde{\tau}_1$
coannihilation or the focus point region.  Additionally a subset of
the points in the $\tilde{\tau}_1$ coannihilation region fulfills the
criterion for the $A/H$-funnel, while some points of the focus point
region fulfill the criterion for $\chi_1^{\pm}$ coannihilation.  No
point in the preferred 2D-$2\sigma$ region fulfills the criterion for
$\tilde{t}_1$ coannihilation, due to relatively large stop masses.

At large $M_0$ and low $M_{1/2}$ a thin strip of our preferred
2D-$2\sigma$ region is excluded at 95\% confidence level by ATLAS jets
plus missing transverse momentum searches requiring exactly one lepton~\cite{Aad:2015mia} or at most one lepton and b-jets~\cite{Aad:2014lra}
in the final state. Therefore an inclusion of these results in the fit
is expected to remove this small part of the focus point region
without changing any conclusion.

Also note that the parameter space for values of $M_0$ larger than
$10$\,TeV was not scanned such that the preferred 2D-$2\sigma$ focus
point region is cut off at this value.  Because the decoupling limit
has already been reached at these large mass scales we do not expect
significant changes in the predicted observable values when going to
larger values of $M_0$.  Hence we also expect the 1D-$1\sigma$ region
to extend to larger values of $M_0$ than visible in
Fig.~\ref{fig:MediumObsSet:PL:M0M12} due to a low sampling density
directly at the $10$\,TeV boundary. For the same reason this cut is
not expected to influence the result of the $p$-value calculation. If
it does it would only lead to an overestimation of the $p$-value.

In the $\tilde{\tau}_1$ coannihilation region negative values of $A_0$
between $-4000$\,GeV and $-1400$\,GeV and moderate values of
$\tan\beta$ between $6$ and $35$ are preferred, while in the focus
point region large positive values of $A_0$ above 3400\,GeV and large
values of $\tan\beta$ above 36 are favored. This can be seen in the
($A_0,\,\tan\beta$)--projection shown in Fig.~\ref{fig:MediumObsSet:PL:A0TanBeta} and in Tab.~\ref{tab:pl:bf}.

\begin{figure}[t]
  \includegraphics[width=0.5\textwidth]{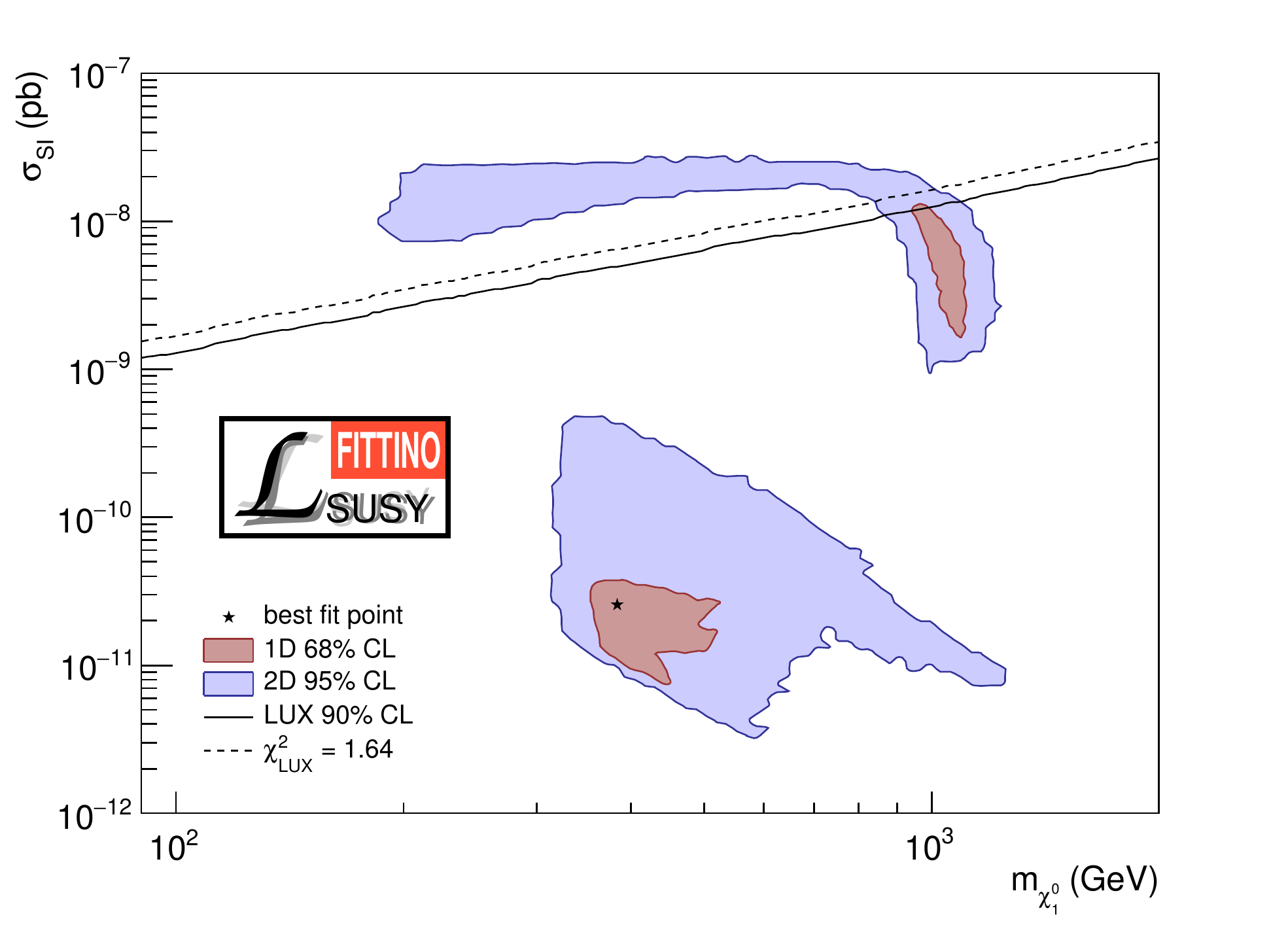}
  \caption{1D$1\sigma$ and 2D$2\sigma$ regions in $m_{\tilde{\chi}_{1}^{0}}$-$\sigma_{SI}$ for the 
  Medium Observable Set. The LUX exclusion is shown in addition. The dashed line indicates a $\chi^2 $ 
  contribution of 1.64 which corresponds to a $90\%$ CL upper limit. This line does not match the LUX 
  exclusion line because we use a theory uncertainty of $50\%$ on $\sigma_{SI}$. }
  \label{fig:MediumObsSet:PL:MassN1SigmaSIChi2}  
\end{figure}

While the $\tilde{\tau}_1$ coannihilation region predicts a spin
independent dark matter--nucleon scattering cross section which is
well below the limit set by the LUX experiment, this measurement has a
significant impact on parts of the focus point region for lightest
supersymmetric particle (LSP) masses
between 200 GeV and 1 TeV, as shown in
Fig.~\ref{fig:MediumObsSet:PL:MassN1SigmaSIChi2}.  The plot also
shows how the additional uncertainty of 50\% on $\sigma_{\textrm{SI}}$
shifts the implemented limit compared to the original limit set by
LUX. It can be seen that future improvements by about 2 orders of
magnitude in the sensitivity of direct detection experiments, as
envisaged \textit{e.g.}  for the future of the XENON~1T
experiment~\cite{2012arXiv1206.6288A}, could at least significantly
reduce the remaining allowed parameter space even taking the
systematic uncertainty into account, or finally discover SUSY dark
matter.

\begin{figure}[t]
  \includegraphics[width=0.5\textwidth]{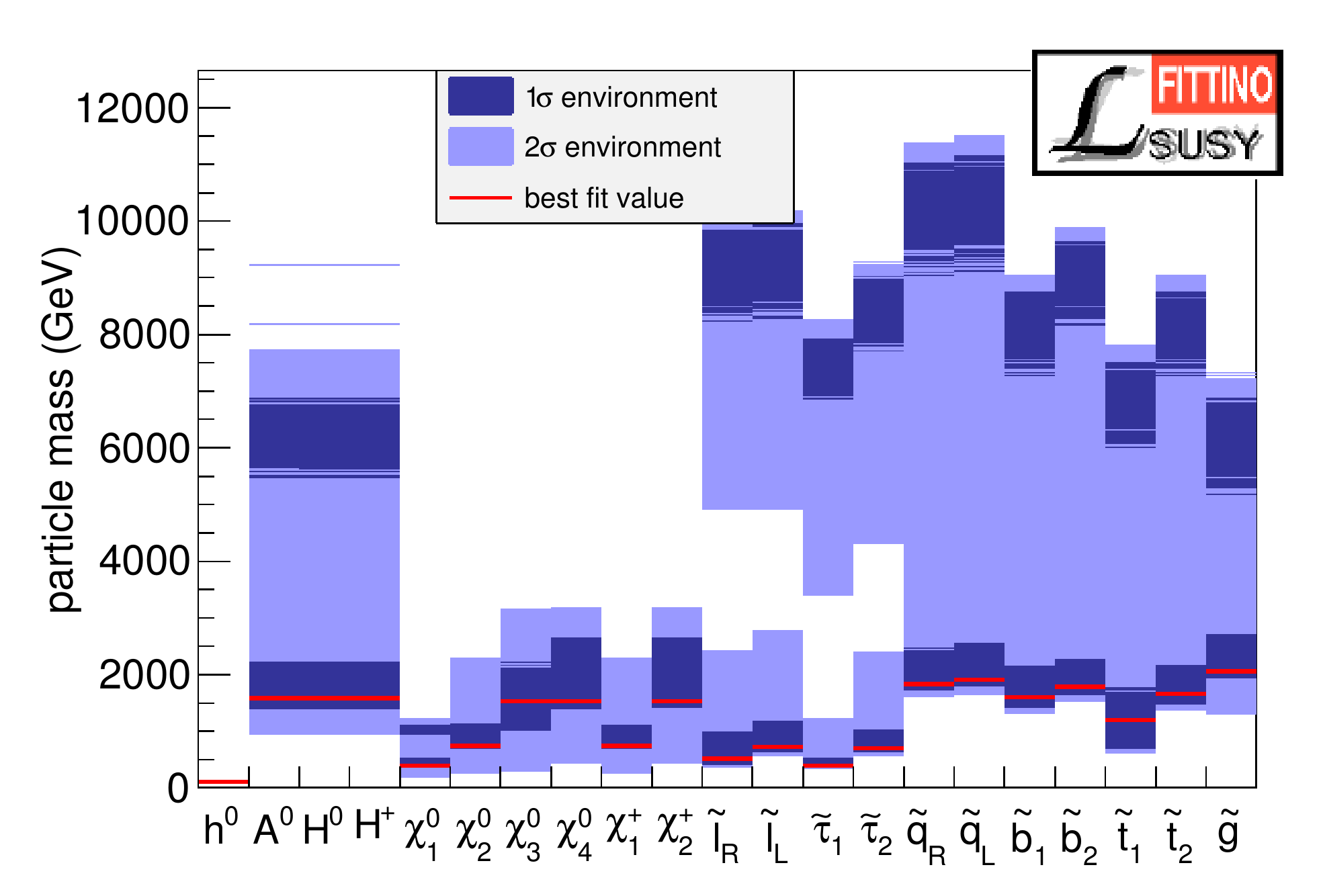}
  \caption{The Higgs and supersymmetric particle mass spectrum as predicted by our fit using the Medium set of Higgs observables.}
  \label{fig:MediumObsSet:PL:mass}  
\end{figure}

The predicted mass spectrum of the Higgs bosons and supersymmetric
particles at the best fit point and in the one-dimensional 1$\sigma$
and 2$\sigma$ regions is shown in
Fig.~\ref{fig:MediumObsSet:PL:mass}.  Due to the relatively shallow
minima of the fit a wide ranges of masses is allowed at $2\sigma$ for
most of the particles. The masses of the coloured superpartners are
predicted to lie above 1.5 TeV, but due to the focus point region also
masses above 10 TeV are allowed at $1\sigma$. Similarly the heavy
Higgs bosons have masses of about 1.5\,TeV at the best fit point, but
masses of about 6\,TeV are preferred in the focus point region. The
sleptons, neutralinos and charginos on the other hand can still have
masses of a few hundred GeV.

\begin{figure}[t]
  \includegraphics[width=0.5\textwidth]{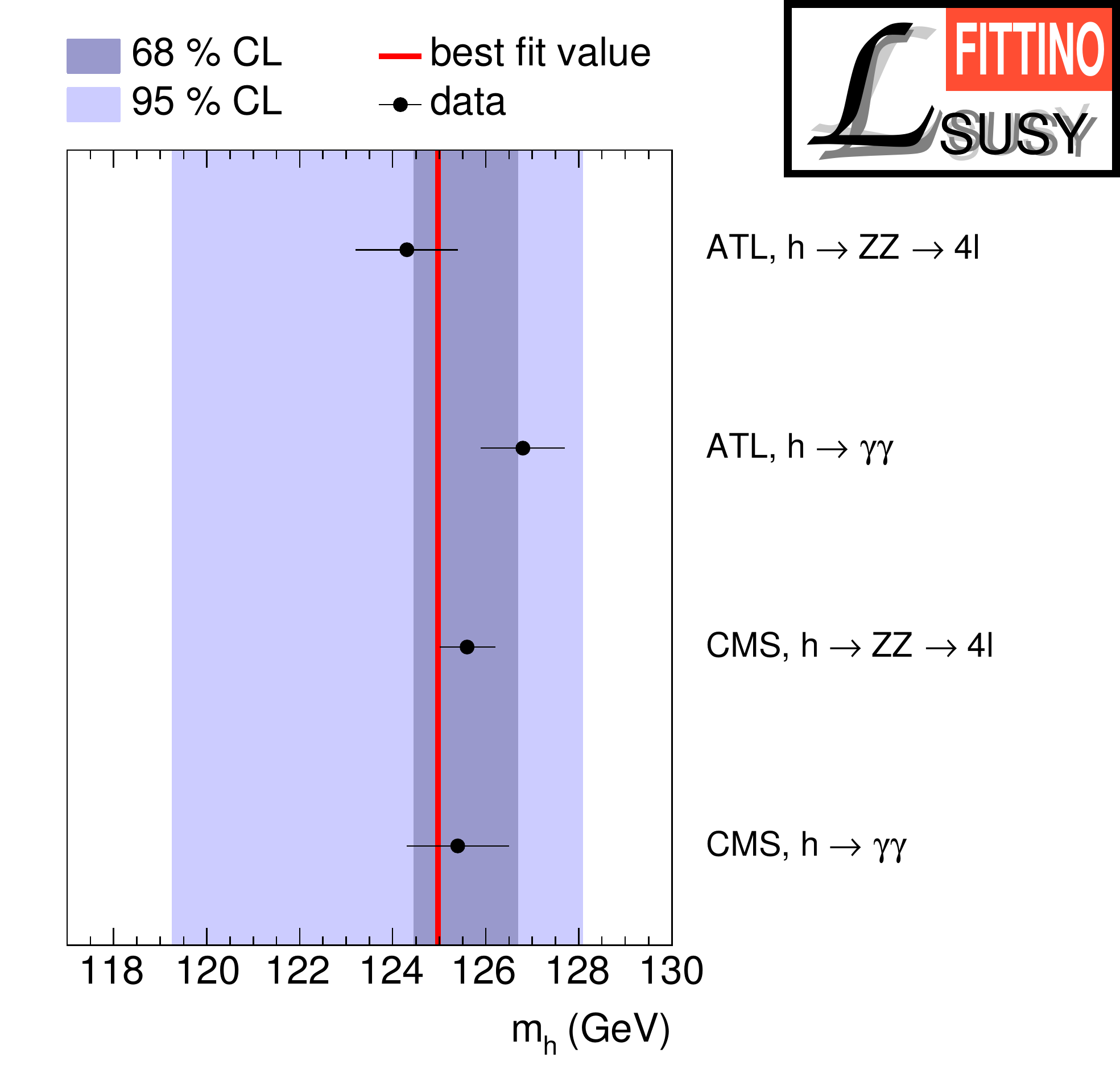}
  \caption{Our predicted mass of the light Higgs boson, together with
    the $1\,\sigma$ and $2\,\sigma$ ranges. The LHC measurements used
    in the fit are shown as well. Note that the correlated theory
    uncertainty of $\Delta m_{h_{theo}}=3$\,GeV is not shown in the
    the plot. The relative smallness of the $68\%$~CL region of the
    fitted mass of $\Delta m_{h_{fit}}=1.1$\,GeV is caused by
    constraints from other observables.}
  \label{fig:MediumObsSet:PL:HiggsMass}
\end{figure}

\begin{figure}[t]
  \includegraphics[width=0.5\textwidth]{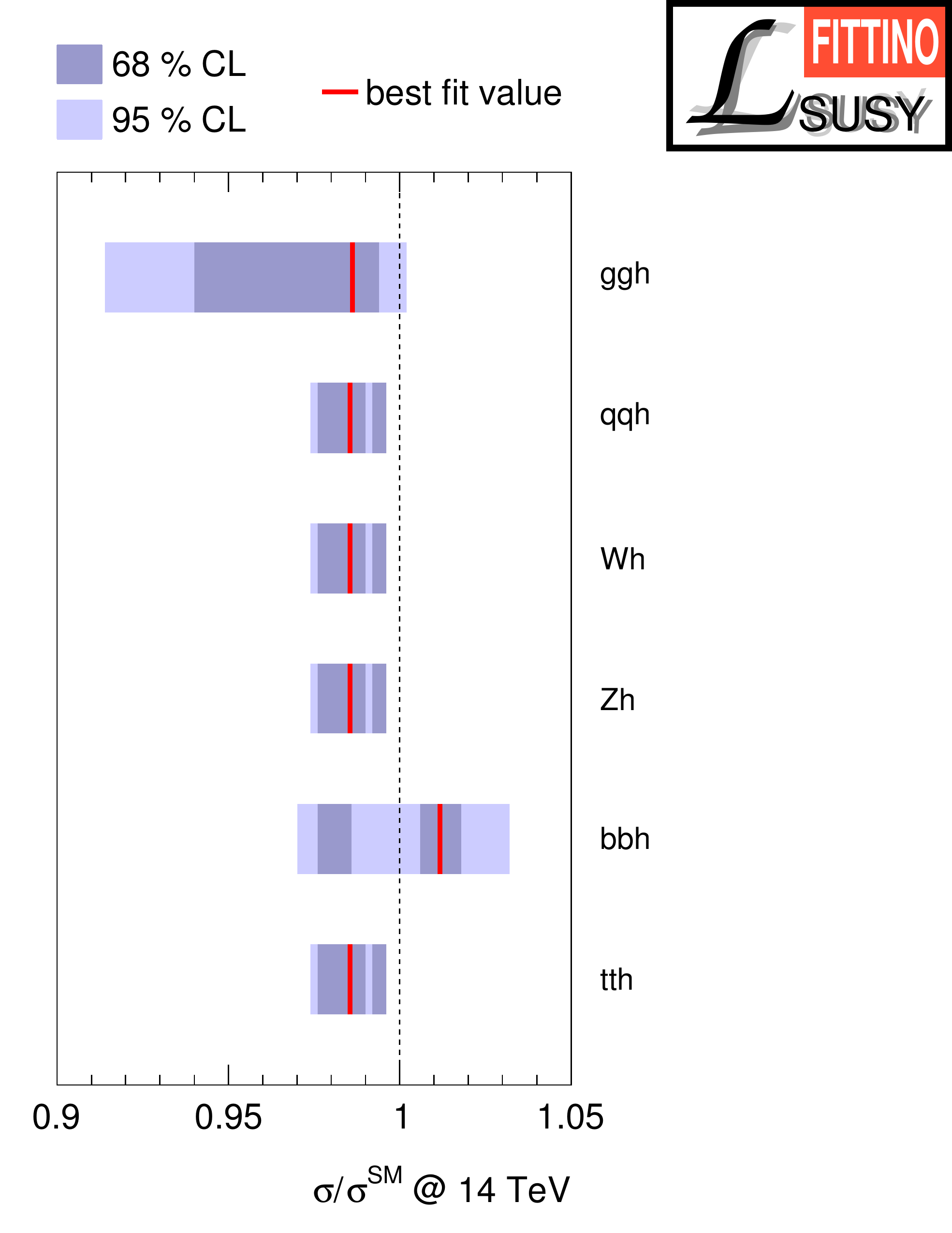}
  \caption{Predicted production cross sections at 14\,TeV of the light Higgs boson relative to the SM value for a Higgs boson with the same mass}
  \label{fig:MediumObsSet:PL:HiggsXS_14TeV}
\end{figure}

\begin{figure}[t]
  \includegraphics[width=0.5\textwidth]{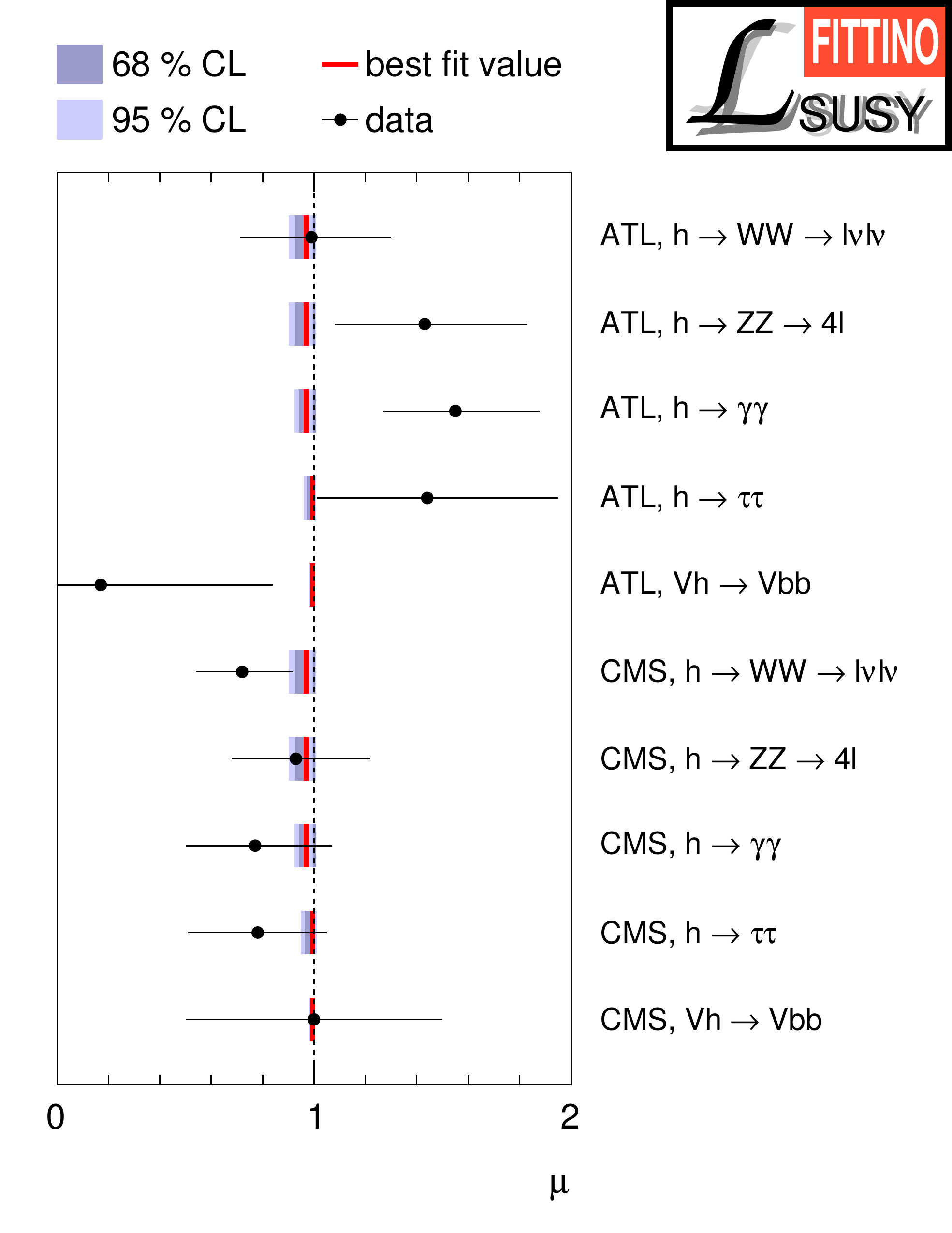}
  \caption{Our predicted $\mu$ values of the light Higgs boson relative to the SM value for a Higgs boson with the same mass. The measurements used in the fit are shown as well.}
  \label{fig:MediumObsSet:HiggsMuData}
\end{figure}

A lightest Higgs boson with a mass as measured by the ATLAS and CMS
collaborations can easily be accommodated, as shown in
Fig.~\ref{fig:MediumObsSet:PL:HiggsMass}. As required by the signal
strength measurements, it is predicted to be
SM-like. Fig.~\ref{fig:MediumObsSet:PL:HiggsXS_14TeV} shows a
comparison of the Higgs production cross sections for different
production mechanisms in p-p collisions at a centre-of-mass energy of
14\,TeV. These contain gluon-fusion (ggh), vector boson fusion (qqh),
associated production (Wh, Zh), and production in assiciation with
heavy quark flavours (bbh, tth). Compared to the SM prediction, the
cMSSM predicts a slightly smaller cross section in all channels except
the bbh channel. The predicted signal strengths $\mu$ in the different
final states for p-p collisions at a centre-of-mass energy of 8 TeV is
also slightly smaller than the SM prediction, as shown in
Fig.~\ref{fig:MediumObsSet:HiggsMuData}. The current precision of
these measurements does, however, not allow for a discrimination
between the SM and the cMSSM based solely on measurements of Higgs
boson properties.

%% file: results_pl_vacuum.tex
\subsection{Vacuum stability}\label{sec:results:pl:vacuum}

The scalar sector of the SM consists of just one complex Higgs doublet. In the cMSSM the scalar sector is dramatically expanded with an extra complex Higgs doublet, as well as the sfermions $\tilde e_{L,R}, \,\tilde \nu_L,\,\tilde u_{L,R},\,\tilde d_{L,R}$ of the first family, and correspondingly of the second and third families. Thus there are 25 complex scalar fields. The corresponding complete scalar potential $V_ {cMSSM}$ is fixed by the five parameters: $(M_0,\,M_{1/2},\,A,\,\tan\beta,\,\mathrm{sgn}(\mu))$. The minimal potential energy of the vacuum is obtained for constant scalar field values everywhere. Given a fixed set of these cMSSM parameters, it is a computational question to determine the minimum of $V_{cMS SM}$. Ideally this minimum should lead to a Higgs vacuum expectation such that SU(2)$_L\times$U(1)$_ Y\to$U(1)$_\mathrm{EM}$, as in the SM. However, it was observed early on in supersymmetric model building, that due to the extended scalar sector, some sfermions could obtain non-vanishing vacuum expectation values (vevs). There could be additional minima of the scalar potential which would break SU(3)$_c$ and/or U(1)$_{\mathrm{EM}}$ and thus colour and/or charge~\cite{Frere:1983ag,Claudson:1983et,Nilles:1982dy,Derendinger:1983bz}. If these minima are energetically higher than the conventional electroweak breaking minimum, then the latter is considered stable. If any of these minima are lower than the conventional minimum, our universe could tunnel into them. If the tunneling time is longer than the age of the Universe of 13.8\,gigayears~\cite{Ade:2013zuv}, we denote our favored vacuum as metastable, otherwise it is unstable. However, this is only a rough categorisation.  Since even if the tunneling time is shorter than the age of the universe, there is a finite probability, that it will have survived until today. When computing this probability, we set a limit of 10\% survival probability.  We wish to explore here the vacuum stability of the preferred parameter ranges of our fits.

The recent observation of the large Higgs boson mass requires within the cMSSM large stop masses and/or a large stop mass splitting. Together with the tuning of the lighter stau mass to favor the stau co-annihilation region (for the low $M_0$ fit region), this typically drives fits to favor a very large value of $|A_{0}|$ relative to $|M_{0}|$, \textit{cf.}  Tab.~\ref{tab:pl:bf}. (For alternative non-cMSSM models with a modified stop sector, see for example~\cite{Auzzi:2012dv,Evans:2012bf,Chamoun:2014eda,Staub:2015aea}.) This is exactly the region, which typically suffers from the SM-like vacuum being only metastable, decaying to a charge- and/or colour-breaking (CCB) minimum of the potential~\cite{Camargo-Molina:2013sta,Blinov:2013fta}.


For the purpose of a fit, in principle a likelihood value for the compatibility of the lifetime of the SM-like vacuum of a particular parameter point with the observation of the age of the Universe should be calculated and should be implemented as a one-sided limit. Unfortunately, the effort required to compute all the minima of the full scalar potential \textit{and} to compute the decay rates for every point in the MCMC \textit{and} to implement this in the likelihood function is beyond present capabilities~\cite{Camargo-Molina:2013sta}.

Effectively, whether or not a parameter point has an unacceptably short lifetime has a binary yes/no answer.  Therefore, as a first step, and in the light of the results of the possible exclusion of the cMSSM in Section~\ref{sec:results:toy}, we overlay our fit result from Section~\ref{sec:results:pl} over a scan of the lifetime of the cMSSM vacuum over the complete parameter space.

\begin{figure}
  \includegraphics[width=0.5\textwidth]{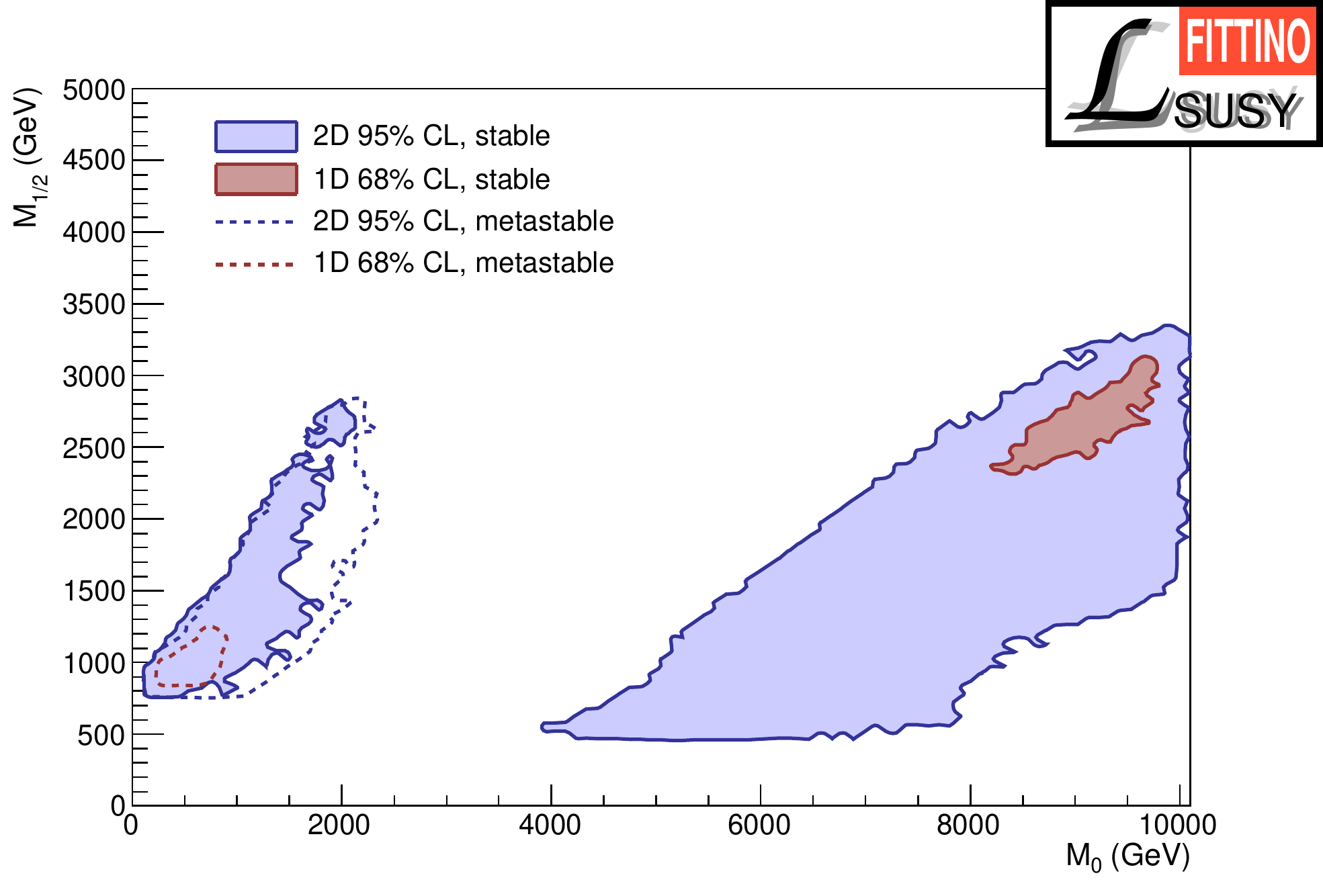}
  \caption{Preferred 1D-1$\sigma$ and 2D-2$\sigma$ regions in $M_0$-$M_{1/2}$ for the Medium Observable Set. The filled areas contain stable points, while the doted lines enclose points which are metastable but still might be very long-lived. The whole preferred 2D-2$\sigma$ focus point region leads to a stable vacuum, while the coannihilation region contains both stable and metastable points. There are no stable points in the preferred 1D-1$\sigma$ coannihilation region.}
  \label{fig:MediumObsSet:PL:CheckVacuum1}  
\end{figure}

\begin{figure}
  \includegraphics[width=0.5\textwidth]{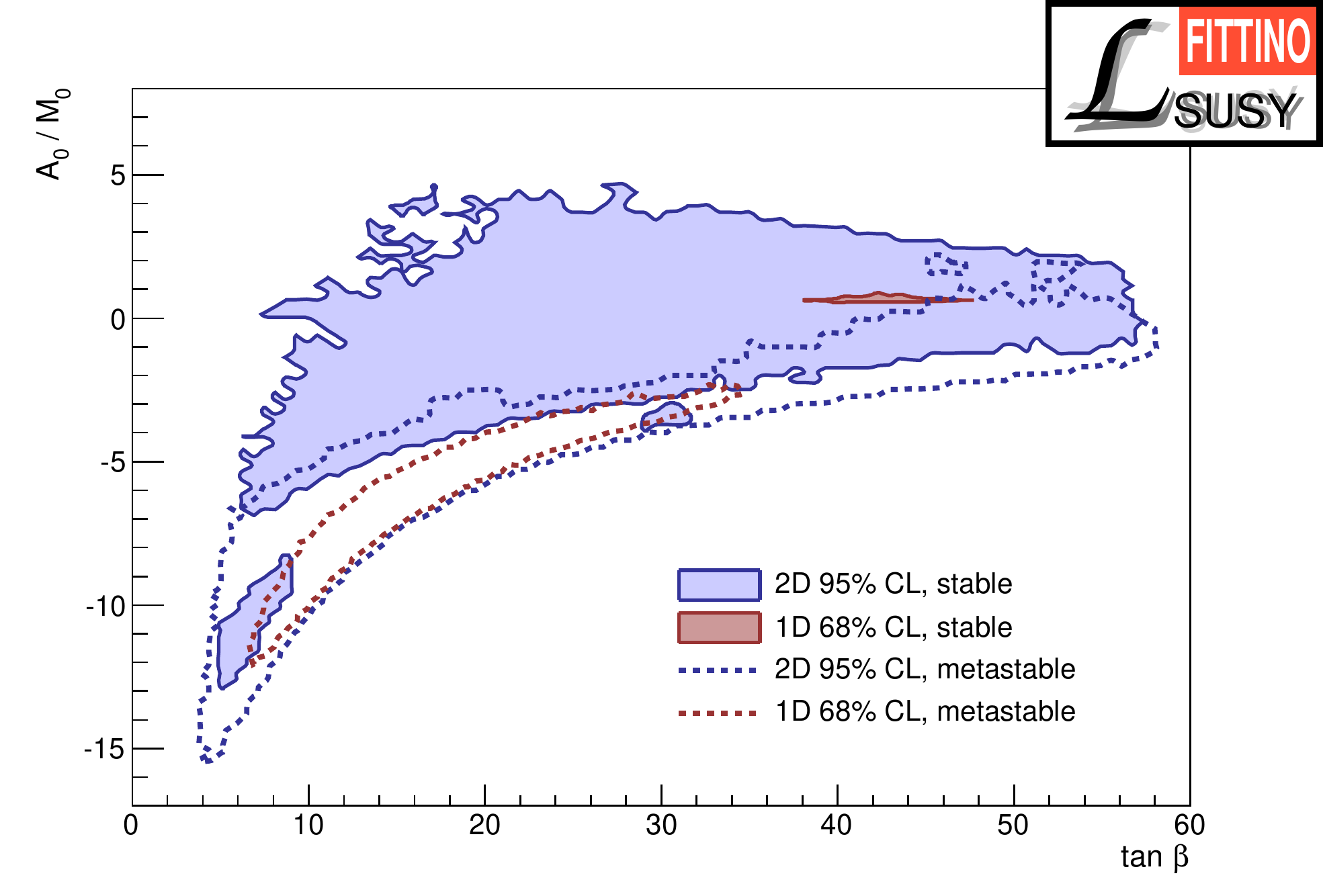}
  \caption{Preferred 1D-1$\sigma$ and 2D-2$\sigma$ regions in $\tan\beta$-$A_0/M_0$ for the Medium Observable Set. The filled areas contain stable points, while the doted lines enclose points which are metastable but still might be very long-lived. Points leading to a metastable vacuum have usually larger negative values of $A_0$ relative to $M_0$ when compared to points with a stable vacuum at the same $\tan\beta$. The part of the 1D-$1\sigma$ region which belongs to the focus point region fulfills $A_0/M_0\sim 0$ and is stable, while the part which belongs to the coannihilation region consists of points with relatively large negative values of $A_0/M_0$ and is metastable.}
  \label{fig:MediumObsSet:PL:CheckVacuum2}  
\end{figure}

The systematic analysis of whether a potential has minima which are deeper than the desired vacuum configuration has been automated in the program \textsc{Vevacious}~\cite{Camargo-Molina:2013qva}. When restricting the analysis to only a most likely relevant subset of the scalar fields of the potential, \textit{i.e.} not the full 25 complex scalar fields, and ignoring the calculation of lifetimes, this code runs sufficiently fast that we are able to present an overlay of which parameter points have CCB minima deeper than the SM-like minimum in Figures~\ref{fig:MediumObsSet:PL:CheckVacuum1} and \ref{fig:MediumObsSet:PL:CheckVacuum2}. However, only the stop and stau fields were allowed to have non-zero values in determining the overlays, in addition to the neutral components of the two complex scalar Higgs doublets. The $\tilde\tau_{L,R},\, \tilde t_{L,R}$ are suspected to have the largest effect~\cite{Camargo-Molina:2013sta}. The computation time when including more scalar fields which are allowed to vary increases exponentially. Thus the more detailed investigations below are restricted to a set of benchmark points. Note that the overlays in Figures~\ref{fig:MediumObsSet:PL:CheckVacuum1} and \ref{fig:MediumObsSet:PL:CheckVacuum2} only show whether metastable vacua might occur at a given point, or whether the vacuum is instable at all. The actual lifetime is not yet considered in this step. See the further considerations below.

There are analytical conditions in the literature for whether MSSM parameter points could have dangerously deep CCB minima, see for example~\cite{Nilles:1982dy,Ibanez:1982qk,AlvarezGaume:1983gj,Claudson:1983et,Derendinger:1983bz,Kounnas:1983td,Ibanez:1984vq,Casas:1995pd}.  These can be checked numerically in a negligible amount of CPU time, while performing a fit. However, these conditions have been explicitly shown to be neither necessary nor sufficient~\cite{Gunion:1987qv}. In particular they have also been shown numerically to be neither necessary nor sufficient for the relevant regions of the cMSSM parameter space which we consider here~\cite{Camargo-Molina:2013sta}.

Since the exact calculation of the lifetime of a metastable SM-like vacuum is so computationally intensive, we unfortunately must restrict this to just the best-fit points of the stau co-annihilation and focus point regions of our the fit, as determined in Section~\ref{sec:results:pl}. As an indicator, though, the extended $\tilde{\tau}_1$ co-annihilation region of the cMSSM investigated in~\cite{Camargo-Molina:2013sta} had SM-like vacuum lifetimes, which were all acceptably long compared to the observed age of the Universe.

The 1D1$\sigma$ best-fit points in Section~\ref{sec:results:pl} where checked for undesired minima, allowing, but not requiring, simultaneously for all the following scalar fields to have non-zero, real VEVs: $H_{d}^{0}, H_{u}^{0}, {\tilde{\tau}}_{L}, {\tilde{\tau}}_{R},{\tilde{\nu}}_{{\mu}L}, {\tilde{b}}_{L}, {\tilde{b}}_{R},{\tilde{t}}_{L}, {\tilde{t}}_{R}$.  The focus point region best-fit point was found to have an absolutely stable SM-like minimum against tunneling to other minima, as no deeper minimum of $V_{cMSSM}$ was found at the 1-loop level.  The SM-like vacuum of the best-fit $\tilde{\tau}_1$ co-annihilation point was found to be metastable, with a deep CCB minimum with non-zero stau and stop VEVs. Furthermore there were unbounded-from-below directions with non-zero values for the $\mu$-sneutrino scalar field in combination with nonzero values for both staus, or both sbottoms, or both chiralities of both staus and sbottoms. This does not bode very well for the absolute best-fit point of our cMSSM fit. However, further effects must be considered.

The parameter space of the MSSM which has directions in field space, where the tree-level potential is unbounded from below was systematically investigated in Ref.~\cite{Casas:1995pd}. We confirmed the persistence of the runaway directions at one loop with \textsc{Vevacious} out to field values of the order of twenty times the renormalisation scale. This is about the limit of trustworthiness of a fixed-order, fixed-renormalisation-scale calculation~\cite{Camargo-Molina:2013qva}. However, this is not quite as alarming as it may seem. The appropriate renormalisation scale for very large field values should be of the order of the field values themselves, and for field values of the order of the GUT scale, the cMSSM soft SUSY-breaking mass-squared parameters by definition are positive. Thus the potential at the GUT scale is bounded from below, as none of the conditions for unbounded-from-below directions given in~\cite{Casas:1995pd} can be satisfied without at least one negative mass-squared parameter. Note, even the Standard Model suffers from a potential which is unbounded from below at a fixed renormalisation scale. Though in the case of the SM it only appears at the one-loop level. Nevertheless, RGEs show that the SM is bounded from below at high energies~\cite{Isidori:2001bm}.

Furthermore, the calculation of a tunneling time \textit{out} of a false minimum does not technically require that the Universe tunnels into a deeper \textit{minimum}. In fact, the state which dominates tunneling is always a vacuum bubble, with a field configuration inside, which classically evolves to the true vacuum after quantum tunneling~\cite{Coleman:1977py,Callan:1977pt}.  Hence the lifetime of the SM-like vacuum of the stau co-annihilation best-fit point could be calculated at one loop even though the potential is unbounded from below at this level. The minimal energy barrier through which the SM-like vacuum of this point can tunnel is associated with a final state with non-zero values for the scalar fields $H_{d}^{0},\, H_{u}^{0},\, {\tilde{\tau}}_{L},\, {\tilde{\tau}}_{R}$, and ${\tilde{\nu}}_{{\mu}L}$. The lifetime was calculated by using the program \textsc{Vevacious} through the program \textsc{CosmoTransitions}~\cite{Wainwright:2011kj} to be roughly $e^{4000} \sim 10^{1700}$ times the age of the Universe. Therefore, we consider the $\tilde{\tau}_1$ coannihilation region best-fit point as effectively stable.

As well as asking whether a metastable vacuum has a lifetime at least as long as the age of the Universe at zero temperature, one can also ask whether the false vacuum would survive a high-temperature period in the early Universe. Such a calculation has been incorporated into \textsc{Vevacious}~\cite{Camargo-Molina:2014pwa}. In addition to the fact that the running of the Lagrangian parameters ensures that the potential is bounded from below at the GUT scale, the effects of non-zero temperature serve to bound the potential from below, as well. In fact the CCB minima of $V_{cMSSM}$ evaluated at the parameters of the stau co-annihilation best-fit point are no longer deeper than the configuration with all zero VEVs, which is assumed to evolve into the SM-like minimum as the Universe cools, for temperatures over about 2300 GeV.  The probability of tunneling into the CCB state integrated over temperatures from 2300 GeV down to 0 GeV was calculated to be roughly $\exp(e^{-2000})$. So while having a non-zero-temperature decay width about $e^{-2000}/e^{-4000} = e^{+2000}$ times larger than the zero-temperature decay width, the SM-like vacuum, or its high-temperature precursor, of the stau co-annihilation best-fit point has a decay probability which is still utterly insignificant.

%% file: results_toy.tex
\subsection{Toy based results}\label{sec:results:toy}

Pseudo datasets have been generated for a total of 7 different minima
based on 6 different observable sets.  For the Medium, Small and
Combined Observable Sets, roughly 1000 sets of pseudo measurements
have been taken into account, as well as for the observable set
without the Higgs rates. For the Medium Observable Set, in addition to
the best fit point, we also study the $p$-value of the local minimum
in the focus point region.  Due to relaxed requirements on the
statistical uncertainty of a $p$-value in the range of
${\cal O}(0.5)$, as compared to ${\cal O}(0.05)$, we use only 125
pseudo datasets for the Large Observable Set. Finally, to study the
importance of (g-2)$_{\mu}$, a total of 500 pseudo datasets have been
generated based on the best fit point for the Medium Observable Set
without (g-2)$_{\mu}$. A summary of all $p$-values with their statistical uncertainties and a comparison to
the naive $p$-value according to the $\chi^{2}$-distribution for
Gaussian distributed variables is shown in Tab.~\ref{tab:pv:pvalues}.

\begin{table*}[t]
    \caption{Summary of $p$-values}
    \label{tab:pv:pvalues}
    \begin{tabular}{c| c| c | c }
    Observable Set & $\chi^2$/ndf & naive $p$-value (\%) & toy $p$-value (\%) \\
    \hline
        Small & 27.1/16& 4.0 & $1.9 \pm 0.4$ \\
            Medium & 30.4/22 & 10.8 & $4.9 \pm 0.7$ \\
    Combined & 17.5/13 & 17.7 & $8.3 \pm 0.8$ \\
\hline
    Medium (Focus Point) & 30.8/22 & 10.0 & $7.8 \pm 0.8$ \\ 
    Medium without (g-2) & 18.1/21 & 64.1  & $51 \pm 3$ \\
    Observable Set without Higgs rates & 15.5/9 & 7.8 & $1.3 \pm 0.4$ \\
    \end{tabular}
\end{table*}

\begin{figure*}[t]
  \begin{center}
    \subfigure[Minimal $\chi^2$ values from toy fits using the \mbox{Medium Observable Set}.]{
      \centering\includegraphics[width=0.45\textwidth]{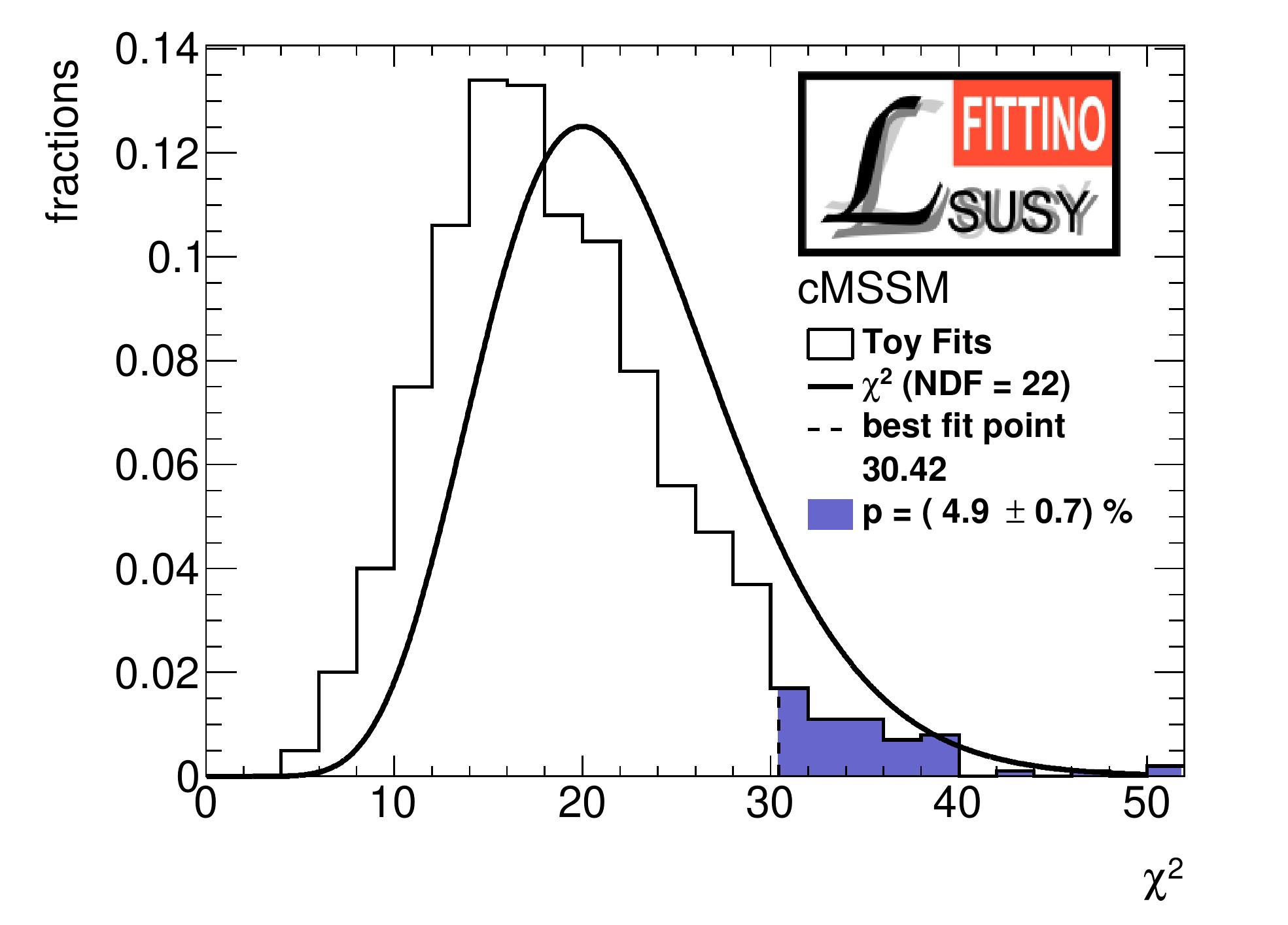}
      \label{fig:MediumObsSet:Toys:Chi2}  
    }
    \subfigure[Minimal $\chi^2$ values from toy fits using the \mbox{Combined Observable Set}.]{
      \centering\includegraphics[width=0.45\textwidth]{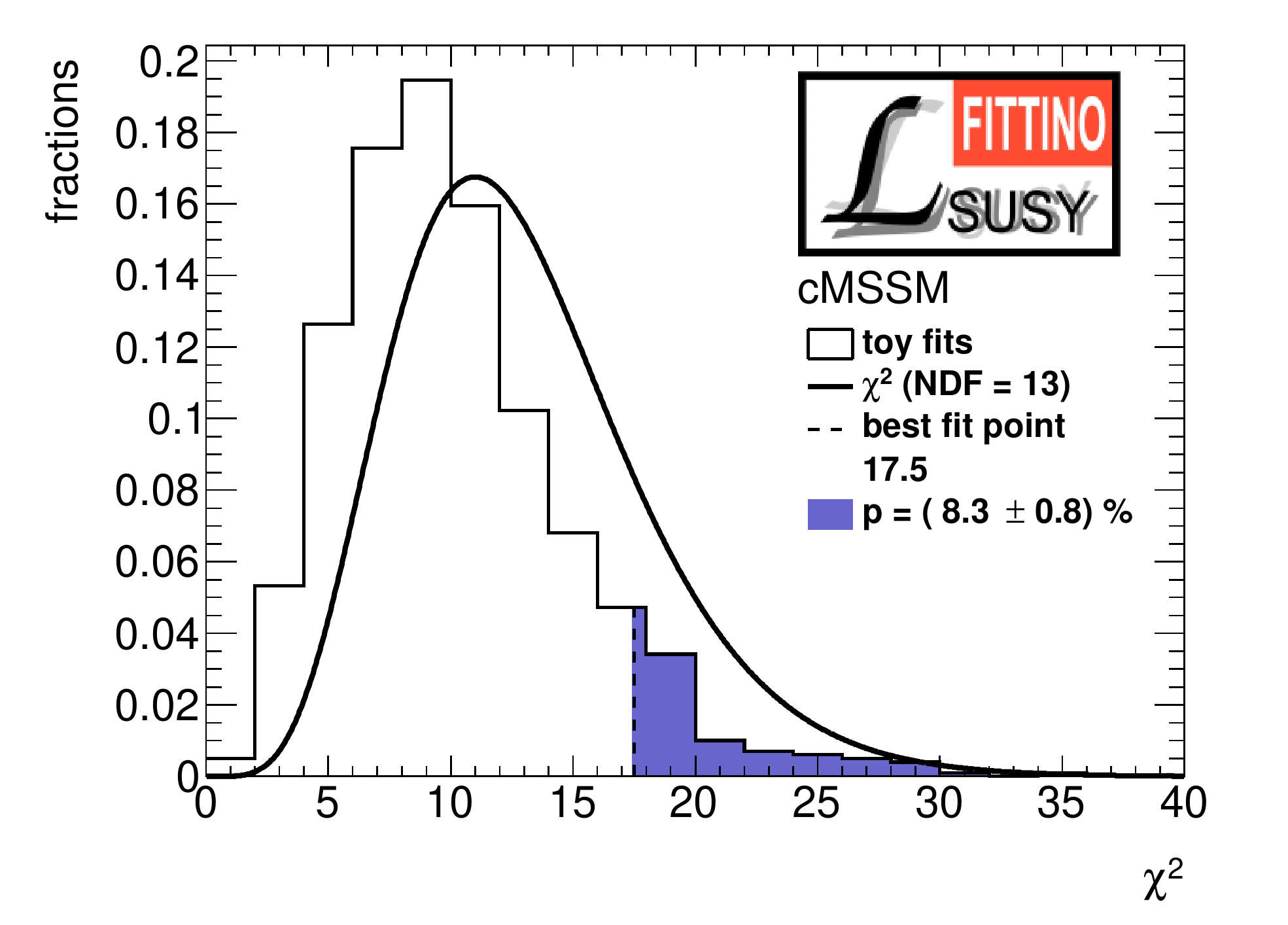}
      \label{fig:CombinedObsSet:Toys:Chi2}  
    }
  \end{center}
  \caption{Distribution of minimal $\chi^2$ values from toy fits using
    two different sets of Higgs observables. A $\chi^2$ distribution
    for Gaussian distributed variables is shown for
    comparison.}\label{fig:Toys:Chi2}
\end{figure*}

\begin{figure*}[t]
  \begin{center}
    \subfigure[Toy fits
    smeared around the global minimum in the
    $\tilde{\tau}$-coannihilation region.]{
      \centering\includegraphics[width=0.45\textwidth]{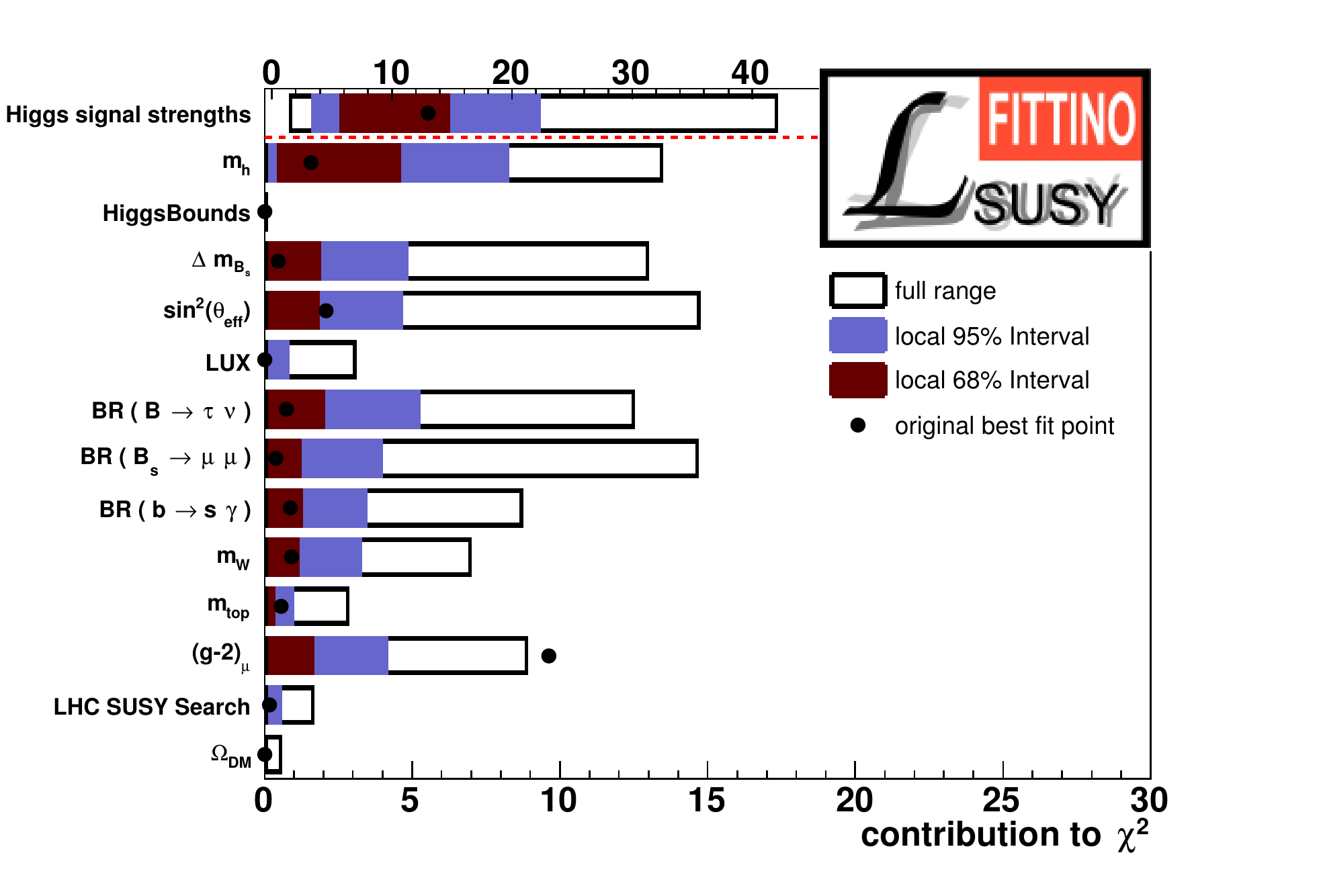}
      \label{fig:MediumObsSet:Toys:Chi2Summary_ToyValues}  
    }
    \subfigure[Toy fits
    smeared around the local minimum in the focus point region.]{
      \centering\includegraphics[width=0.45\textwidth]{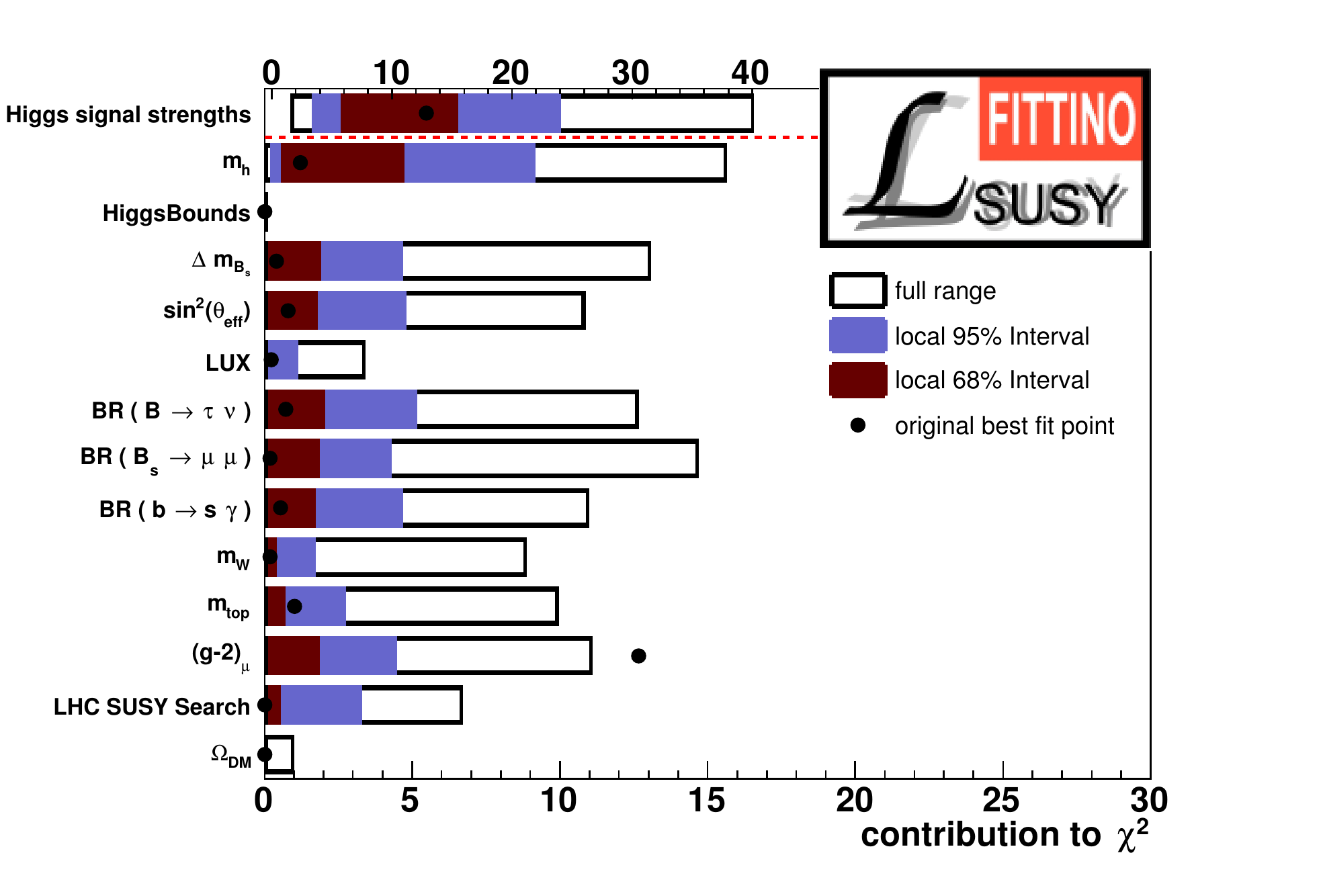}
      \label{fig:MediumObsSet_FocusPoint:Toys:Chi2Summary_ToyValues}  
    }
  \end{center}
  \caption{Individual $\chi^2$ contributions of all 
    observables/observable sets at the best fit points of the toy 
    fits using the Medium Set of Higgs observables with observables 
    smeared around the global and the local minimum of the observed
    $\chi^2$ contour. The 
    predicted measurements at the best fit points of the individual
    pseudo data fits are used to derive the local CL intervals shown
    in the plots. These are compared with the individual $\chi^2$
    contribution of each observable at the global or local minimum. Note the 
    different scale shown on the top which is used for HiggsSignals,
    which contains 14 observables. Also note that $m_{h}$ contains 
    contributions from $4$ measurements for this observable set.}\label{fig:Toys:Chi2Summary_ToyValues}
\end{figure*}

Figure~\ref{fig:MediumObsSet:Toys:Chi2} shows the
$\chi^{2}$-distribution for the best fit point of the Medium
Observable Set, from which we derive a $p$-value of ($4.9\pm0.7$)\%.
As a comparison we also show the $\chi^{2}$-distribution for the pseudo fits 
using the Combined Observable Set in figure \ref{fig:CombinedObsSet:Toys:Chi2}. 
Both distributions are significantly shifted towards smaller $\chi^{2}$-values compared to the corresponding $\chi^{2}$-distributions for Gaussian
distributed variables.  Several observables are responsible for the
large deviation between the two distributions, as shown in
Fig.~\ref{fig:MediumObsSet:Toys:Chi2Summary_ToyValues}, where the
individual contributions of all observables to the minimum $\chi^{2}$
of all pseudo best fit points are plotted.

First, \textsc{HiggsBounds} does not contribute significantly to the
$\chi^{2}$ at any of the pseudo best-fit points, which is also the
case for the original fit. The reason for this is, that for the
majority of tested points, the $\chi^{2}$ contribution from
\textsc{HiggsBounds} reflects the amount of violation of the LEP limit
on the lightest Higgs boson mass by the model. Since the measurements
of the Higgs mass by ATLAS and CMS lie significantly above this limit,
it is extremely unlikely that in one of the pseudo datasets the Higgs
mass is rolled such that the best fit point would receive a $\chi^{2}$
penalty due to the LEP limit. This effectively eliminates one degree
of freedom from the fit. In addition, the predicted masses of $A,H$
and $H^{\pm}$ lie in the decoupled regime of the allowed cMSSM
parameter space. Thus there are no contributions from heavy Higgs or
charged Higgs searches as implemented in \textsc{HiggsBounds}.

The same effect is observed slightly less pronounced for the LHC and
LUX limits, where the best fit points are much closer to the
respective limits than in the case of \textsc{HiggsBounds}. Finally we
observe that for each pseudo dataset the cMSSM can very well describe
the pseudo measurement of the dark matter relic density, which further
reduced the effective number of degrees of freedom.

\begin{figure*}[t]
  \begin{center}
    \subfigure[Anomalous magnetic moment of the muon.]{
      \centering\includegraphics[width=0.45\textwidth]{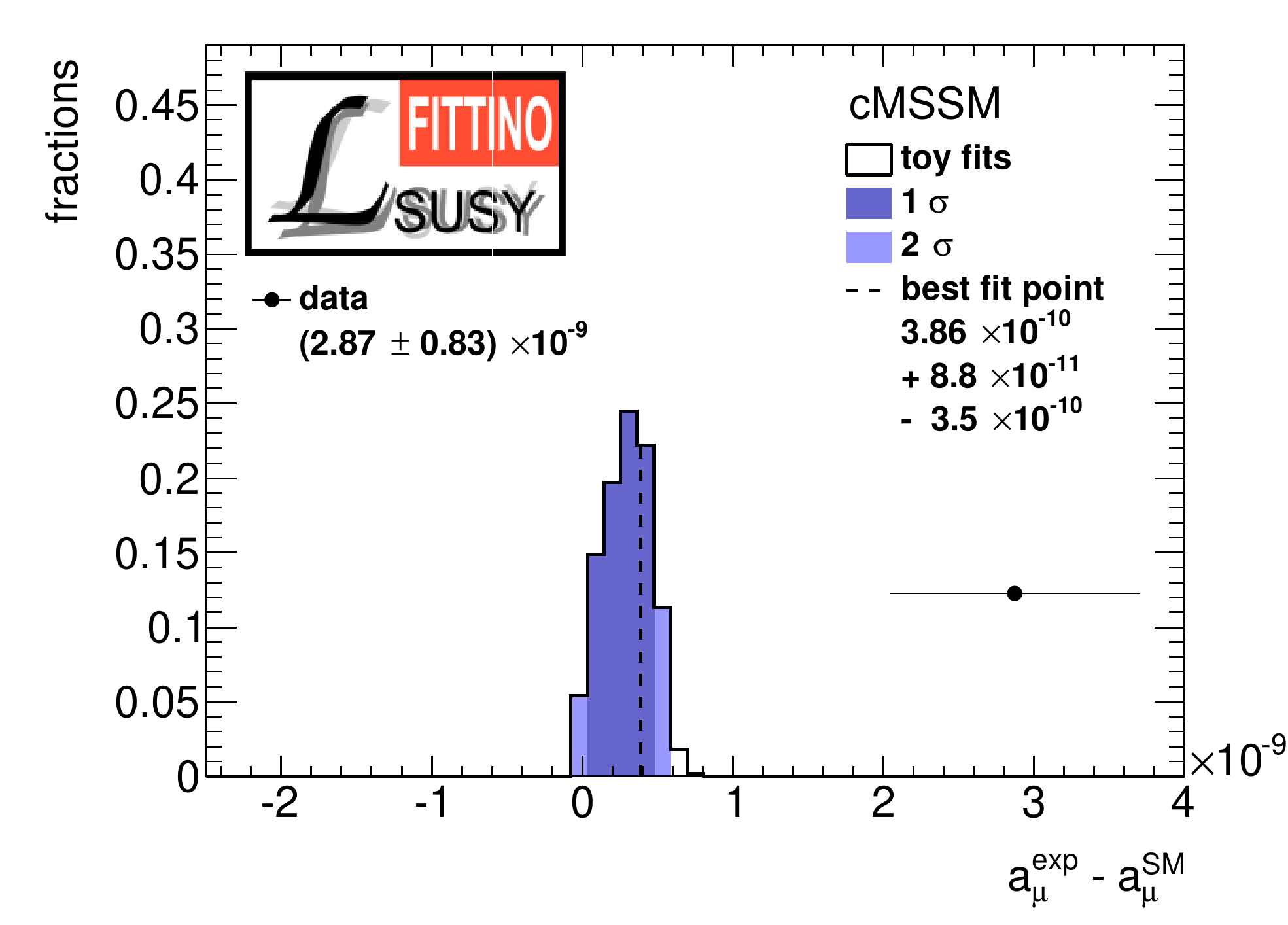}
      \label{fig:MediumObsSet:Toys:GMin2Muon}  
    }
    \subfigure[Dark matter relic density.]{
      \centering\includegraphics[width=0.45\textwidth]{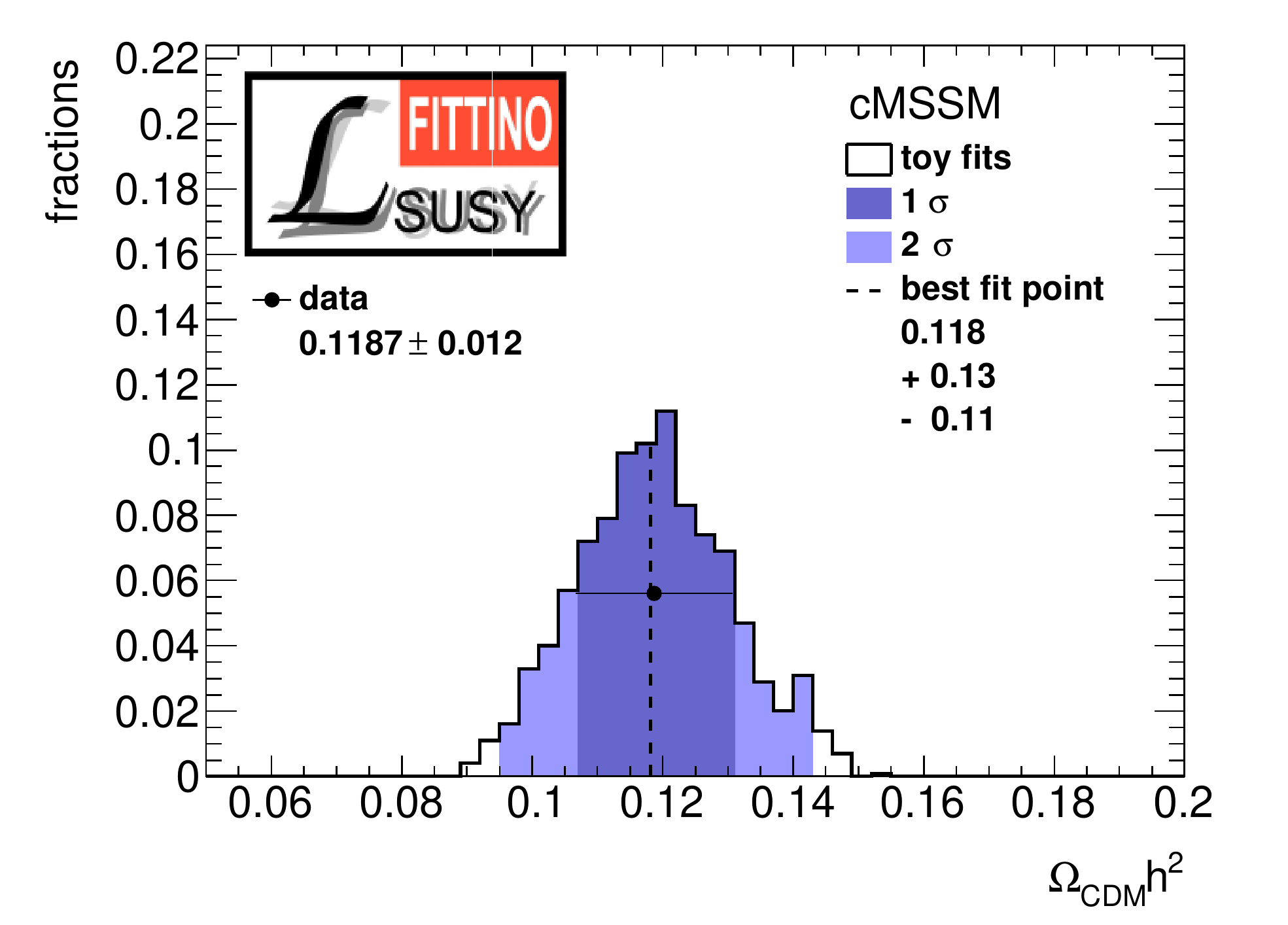}
      \label{fig:MediumObsSet:Toys:Omega} 
    } 
  \end{center}
  \caption{Distribution of the predictions of the best fit points of
    the pseudo data fits for two
    different observables used in the fit, compared with the
    respective measurements.}\label{fig:MediumObsSet:Toys:Observables}
\end{figure*}

Figure~\ref{fig:MediumObsSet:Toys:Chi2Summary_ToyValues} also shows
that the level of disagreement between measurement and prediction for
$(g-2)_{\mu}$ is smaller in each single pseudo dataset than in the
original fit with the real dataset. The 1-dimensional distribution of
the pseudo best fit values of $(g-2)_ {\mu}$ is shown in
Fig.~\ref{fig:MediumObsSet:Toys:GMin2Muon}. The figure shows that
under the assumption of our best fit point, not a single pseudo
dataset would yield a prediction of $(g-2)_{\mu}$ that is consistent
with the actual measurement. As a comparison,
Fig.~\ref{fig:MediumObsSet:Toys:Omega} shows the 1-dimensional
distribution for the dark matter relic density, where the actual
measurement can well be accommodated in any of the pseudo best fit
scenarios. To further study the impact of $(g-2)_{\mu}$ on the
$p$-value, we repeat the toy fits without this observable and get a
$p$-value of $(51 \pm 3)\%$.  This shows that the relatively low
$p$-value for our baseline fit is mainly due to the incompatibility of
the $(g-2)_{\mu}$ measurement with large sparticle masses, which are
however required by the LHC results.

\begin{figure}[t]
   \includegraphics[width=0.5\textwidth]{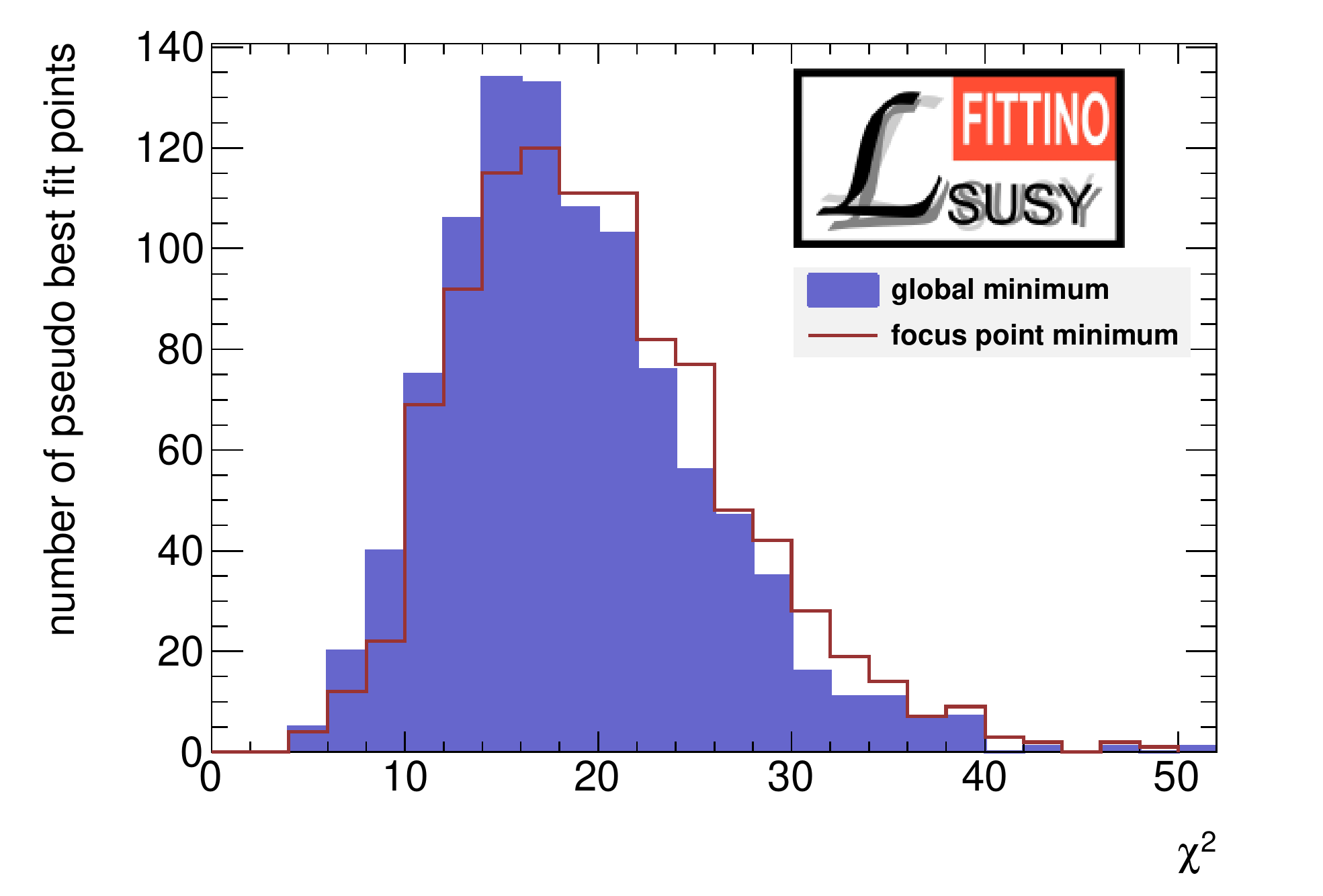}
   \caption{Comparison of the $\chi^{2}$ distributions obtained from toy fits using the global minimum and the local minimum in the focus point region of the Medium Observable Set.}
  \label{fig:MediumObsSet:Toys:Chi2Comparison}  
\end{figure}

Interestingly, under the assumption that the minimum in the focus
point region is the true description of nature, we get a slightly
better $p$-value (7.8\%) than we get with the actual best fit
point. Figure~\ref{fig:MediumObsSet_FocusPoint:Toys:Chi2Summary_ToyValues}
shows the individual contributions to the pseudo best fit $\chi^{2}$
at the pseudo best fit points for the toy fits performed around the
local minimum in the focus point region. There are two variables with
higher average contributions compared to the global minimum: $m_{top}$
and the LHC SUSY search. In particular for the LHC SUSY search, the
LHC contribution to the total $\chi^2$ is, on average, significantly
higher than for the pseudo best fit points for the global minimum. The
number of expected signal events for the minimum in the focus point
region is 0, while it is $> 0$ for the global minimum. Pseudo best fit
points with smaller values of the mass parameters, in particular
pseudo best fit points in the $\tilde{\tau}$-coannihilation region,
tend to predict an expected number of signal events larger than
zero. Since for the pseudo measurements based on the minimum in the
focus point region an expectation of 0 is assumed, this naturally
leads to a larger $\chi^{2}$ contribution from the ATLAS 0-$\ell$
analysis. The effect on the distribution of the total $\chi^{2}$ is
shown in figure \ref{fig:MediumObsSet:Toys:Chi2Comparison}.  Another
reason might be that the focus point region is sampled more coarsely
than the region around the global minimum. This increases the
probability that the fit of the pseudo dataset misses the actual best
fit point, due to our procedure of using only the points in the
original MCMC.  This effect should however only play a minor role,
since the parameter space is still finely scanned and only a
negligible fraction of scan points are chosen numerous times as best
fit points in the pseudo data fits.

To further verify that this effect is not only caused by the coarser sampling in the 
focus point region, we performed another set of 500 pseudo fits based on the 
global minimum, reducing the point density in the $\tilde{\tau}$-coannihilation region 
such that it corresponds to the point density around the local minimum in the focus point region. 
We find that the resulting $\chi^{2}$ distribution is slightly shifted with respect to the 
$\chi^{2}$ distribution we get from the full MCMC. The shift is, however, too small to explain the 
difference between $p$-values we find for the global minimum and the local minimum in the focus point 
region.

As an additional test, we investigate a simple toy model with only Gaussian observables and 
a single one-sided limit corresponding to the LHC SUSY search we use in our fit of the cMSSM. 
Also in this very simple model we find significantly different $\chi^{2}$ distributions for 
fits based on points in a region with/without a significant signal expectation for the counting 
experiment. We thus conclude that the true $p$-value for the local minimum in the focus point region 
is in fact higher than the true $p$-value for the global minimum of our fit.

In order to ensure that there are no more points with a higher $\chi^{2}$ and a higher $p$-value than the local 
minimum in the focus point region, we generate 200 pseudo datasets for two more points in the focus point region. 
The two points are the points with the highest/lowest $M_{0}$ in the local 1$\sigma$ region around the focus point minimum. 
We find that the $\chi^{2}$ distributions we get from these pseudo datasets are in good agreement with each other and also 
with the $\chi^{2}$ distribution derived from the pseudo experiments around the focus point minimum, and hence conclude that 
the local minimum in the focus point region is the point with the highest $p$-value in the cMSSM.

To study the impact of the Higgs rates on the $p$-value, and in order
to compare to the observable sets used by other fit collaborations,
which exclude the Higgs rate measurements from the fit on the basis
that \textit{in the decoupling regime} they do not play a vital role,
we perform toy fits for the observable set without Higgs rates and
derive a $p$-value of $(1.3\pm 0.4)\%$. In the decoupling limit, the
cMSSM predictions for the Higgs rates are very close to the SM, so
that the LHC is not able to distinguish between the two models based
on Higgs rates measurements (see
Fig.~\ref{fig:MediumObsSet:HiggsMuData}). Because of the overall
good agreement between the Higgs rate measurements and the SM
prediction, the inclusion of the Higgs rates in the fit improves the
fit quality despite some tension between the ATLAS and CMS
measurements.

As discussed in Section~\ref{sec:methods}, it is important to
understand the impact of the parametrisation of the measurements on
the $p$-value. To do so, we compare our baseline fit with two more
extreme choices.  First, we use the Small Observable Set which
combines $h\to\gamma\gamma$, $h \to ZZ$, and $h \to WW$ measurements
but keeps ATLAS and CMS measurements separately.  We use this choice
because an official ATLAS combination is available.  The equivalent
corresponding CMS combination is produced independently by us.  Using
this observable set we get a $p$-value of $(1.9\pm 0.4)\%$.  Here the
cMSSM receives a $\chi^2$ penalty from the trend of the 
ATLAS signal strength measurements to values $\mu\geq1$ and of the CMS
measurements towards $\mu\leq1$ in the three $h \to VV$ channels.

As a cross-check, we employ the Large Observable Set, which contains  
all available sub-channel measurements separately. Using this observable  
set, we get a $p$-value from the pseudo data fits of  
$(41.6 \pm 4.4)\%$. As observed in Section~\ref{sec:results:pl}, the
Large Observable Set yields the same preferred parameter region as the
Small, Medium and Combined Observable sets. Yet, its $p$-value from
the pseudo data fits significantly differs.
 
To explain this interesting result we consider a simplified example:
For $i=1,\dots,N$, let $x_i$ be Gaussian measurements with
uncertainties $\sigma_i$ and corresponding model predictions
$a_i(\mathbf{P})$ for a given parameter point $\mathbf{P}$. We assume
that the measurements from $x_n$ to $x_N$ are uncombined measurements
of the same observable; then $a_i=a_n$ for all $i\geq n$.  There are
now two obvious possibilities to compare measurements and predictions:
\begin{itemize}
\item We can compare each of the individual measurements with 
 the corresponding model predictions by calculating
 \begin{equation*}
 \chi^2_{\mathrm{split}} = \sum_{i=1}^{N} \left(\frac{x_i -
     a_i}{\sigma_i} \right)^2.
\end{equation*}
This would correspond to an approach where the model is confronted
with all avaialable observables, irrespective of the question whether
they measure independent quantities in the model or not. One example
for such a situation would be the Large Observable Set of Higgs signal
strength measurements, where several observables measure different
detector effects, but the same physics.
\item We can first combine the measurements $x_i$, $i\geq n$ to a
  measurement $\bar{x}$ which minimises the function
\begin{equation}
f(x)=\sum_{i=n}^{N}\left( \frac{x_i-x}{\sigma_i} \right)^2
\end{equation}
and has an uncertainty of $\bar{\sigma}$ and then use this combination
to calculate
\begin{equation}
\chi^2_\mathrm{combined}=\sum_{i=1}^{n-1}\left(\frac{x_i-a_i}{\sigma_i} \right)^2 + \left(\frac{\bar{x}-a_n}{\bar{\sigma}}\right)^2. 
\end{equation}
This situation now corresponds to first calculating one physically
meaningful quantity (e.g.{} a common signal strength for
$h\to\gamma\gamma$ in all VBF categories, and all $gg\to h$
categories) and only then to confront the model to the combined
measurement.
\end{itemize}
Plugging in the explicit expressions for ($\bar{x}, \bar{\sigma}$),
using $1/\bar{\sigma}^2=\sum_{i=n}^{N}(1/\sigma_i^2)$ and defining $\chi^2_{\mathrm{data}}=f(\bar{x})$ one finds
\begin{equation}\label{eq:relationofchi2}
\chi^2_\mathrm{combined}=\chi^2_\mathrm{split}-\chi^2_\mathrm{data}.
\end{equation}

Hence doing the combination of the measurements before the fit is
equivalent to using a $\chi^2$-difference which in turn is equivalent
to the usage of a log-likelihood ratio. The numerator of this ratio is
given by the likelihood $\mathcal{L}_{\mathrm{model}}$ of the model
under study, \textit{e.g.}  the cMSSM. The denominator is given by the
maximum of a phenomenological likelihood $\mathcal{L}_\mathrm{pheno}$
which depends directly on the model predictions $a_i$. These possess an
expression as functions $a_i(\mathbf{P})$ of the model parameters
$\mathbf{P}$ of
$\mathcal{L}_\mathrm{model}$. Note that in
$\mathcal{L}_\mathrm{pheno}$ however, the $a_i$ are treated as $n$ 
independent parameters. We now identify $\chi^2_\mathrm{split}\equiv
-2\ln \mathcal{L}_{\mathrm{model}}$ and $\chi^2_\mathrm{data}\equiv
-2\ln \mathcal{L}_\mathrm{pheno}$. When inserting $a_i(\mathbf{P})$,
one is guaranteed to find
\begin{equation}
 \mathcal{L}_\mathrm{pheno}(a_1(\mathbf{P}),\dots, a_n(\mathbf{P})) = \mathcal{L}_{\mathrm{model}}(\mathbf{P}). 
 \end{equation}
Using these symbols, $\chi^2_{\mathrm{combined}}$ can be written as
\begin{equation}\label{eq:LLR}
\chi^2_{\mathrm{combined}}= -2\ln \frac{\mathcal{L_{\mathrm{model}}(\mathbf{P})}}{\mathcal{L}_
\mathrm{pheno}(\hat{a}_1, \dots, \hat{a}_n)}\,,
\end{equation}
where $\hat{a}_1,\dots,\hat{a}_n$ maximise
$\mathcal{L}_{\mathrm{pheno}}$.  Note that in this formulation the
model predictions $a_i$ do not necessarily need to correspond directly
to measurements used in the fit, as it is the case for our
example. For instance the model predictions $a_i$ might contain cross
sections and branching ratios which are constrained by rate
measurements.

Using $\mathrm{ndf}_\mathrm{split}=N$, $\mathrm{ndf}_\mathrm{combined}=n$ and  $\mathrm{ndf}_\mathrm{data}=N-n$ Eq.~(\ref{eq:relationofchi2}) implies  
\begin{equation}
\frac{\chi^2_{\mathrm{split}}}{\mathrm{ndf_{split}}} 
= \frac{\chi^2_\mathrm{combined}}{\mathrm{ndf_{combined}}+N-n}
+ \frac{\chi^2_\mathrm{data}}{\mathrm{ndf_{data}}+n}.
\end{equation}
The more uncombined measurements are used, the larger $N-n$ gets and
the less the $p$-value depends on the first term on the right hand
side, which measures the agreement between data and model. At the same
time, the $p$-value depends more on the second term on the right hand
which measures the agreement within the data.  Especially, for $n$
fixed and $N\to\infty$:
\begin{equation}
\frac{\chi^2_{\mathrm{split}}}{\mathrm{ndf_{split}}} = \frac{\chi^2_{\mathrm{data}}}{\mathrm{ndf_{data}}}\,.
\end{equation}

\begin{figure}[t]
\includegraphics[width=0.5\textwidth]{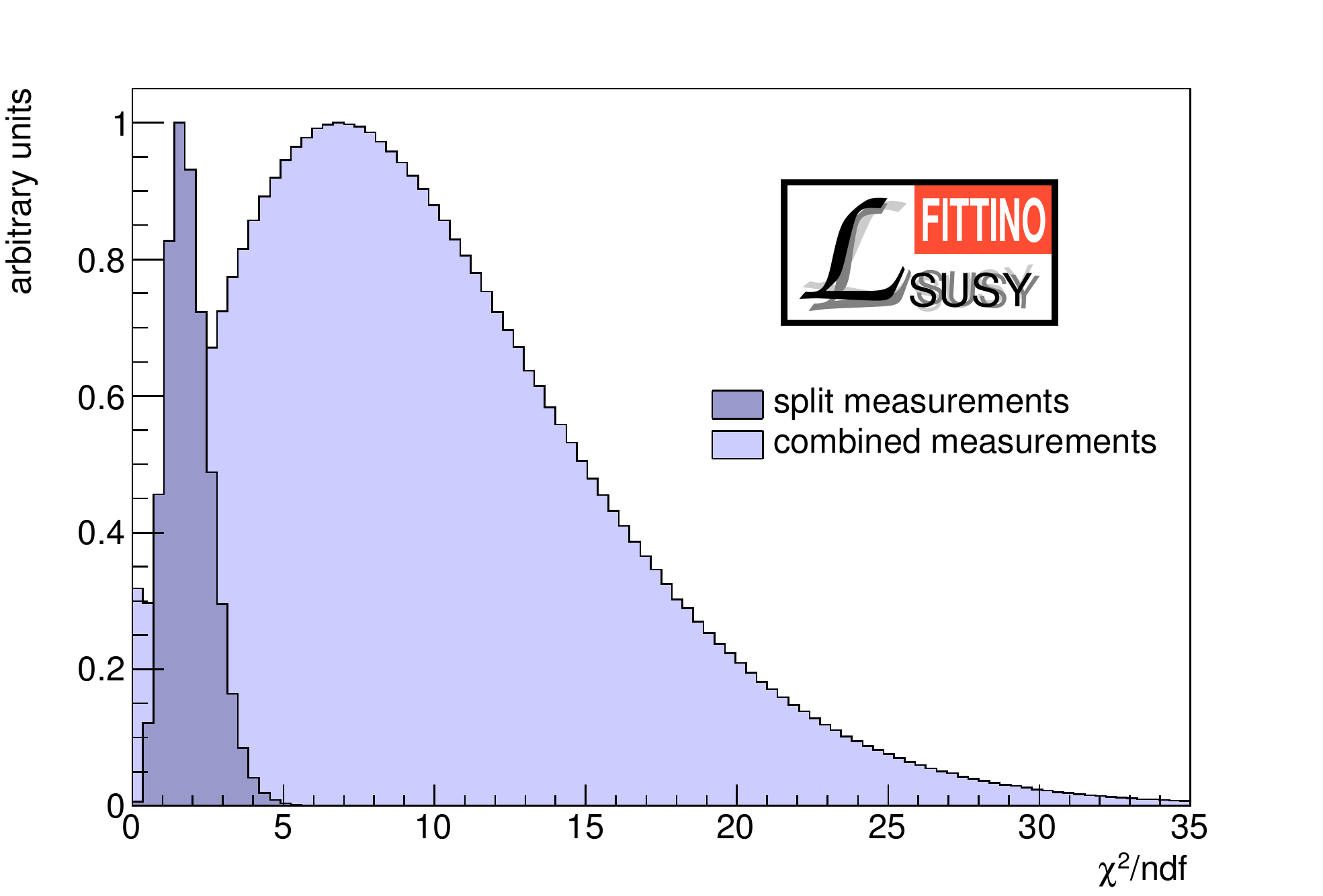}
\caption{Numerical example showing the distribution of $\chi^2/\mathrm{ndf}$ using combined and split measurements. Using split measurements smaller values of $\chi^2/\mathrm{ndf}$ are obtained. Because in this example all measurements are Gaussian, this is equivalent to larger $p$-values. We call the effect of obtaining larger $p$-values when using split measurements   the dilution of the $p$-value.}
\label{fig:toyexample}
\end{figure}

Since in the case of purely statistical fluctuations of the split
measurements around the combined value the agreement
\textit{within} the data is unlikely to be poor, the expectation is
\begin{equation}
\frac{\chi^2_{\mathrm{data}}}{\mathrm{ndf_{data}}}\approx 1
\end{equation}
even if there was a deviation between the model predictions and the
\textit{physical} combined observables. 
So most of the time the $p$-value will get larger when uncombined
measurements are used, hiding deviations between model and data.  As a
numerical example Fig.~\ref{fig:toyexample} shows toy distributions
of
$\frac{\chi^2_{\mathrm{combined}}}{\mathrm{ndf_{combined}}}$ and
$\frac{\chi^2_{\mathrm{split}}}{\mathrm{ndf_{split}}}$ for one
observable (n=1), ten measurements (N=10) and a $3\sigma$ deviation
between the true value and the model prediction.  We call this effect
\textit{dilution of the $p$-value}. It explains the large $p$-value
for the Large Observable Set by the overall good agreement between the
uncombined measurements.

On the other hand if there is some tension within the data, which
might in this hypothetical example be caused purely by statistical or
experimental effects, the ``innocent'' model is punished for these
internal inconsistencies of the data. This is observed here for the
Medium Observable Set and Small Observable Set.  Hence, and in order
to incorporate our assumption that ATLAS and CMS measured the same
Higgs boson, we produce our own combination of corresponding ATLAS and
CMS Higgs mass and rate measurements.  We also assume that custodial
symmetry is preserved but do not assume that $h\to \gamma\gamma$ is
connected to $h\to WW$ and $h\to ZZ$ as in the official ATLAS
combination used in Small Observable Set.  We call the resulting
observable set Combined Observable Set.  Note that for simplicity we
also combine channels for which the cMSSM model predictions might
differ due to different efficiencies for the different Higgs
production channels.  This could be improved in a more rigorous
treatment.  For instance the $\chi^2$ could be defined by
Eq.~(\ref{eq:LLR}) using a likelihood $\mathcal{L}_\mathrm{pheno}$
which contains both the different Higgs production cross sections and
the different Higgs branching ratios as free parameters $a_i$.


Using the Combined Observable Set we get a $p$-value of
$(8.3\pm 0.8)\%$, which is significantly smaller than the diluted
$p$-value of $(41.6\pm 4.4)\%$ for the Large Observable Set. The good
agreement within the data now shows up in a small
$\chi^2/\mathrm{ndf}$ of $68.1/65$ for the combination but no longer
affects the $p$-value of the model fit. On the other hand the
$p$-value for the Combined Observable Set is larger than the one for
the Medium Observable Set, because the tension between the ATLAS and
CMS measurements is not included.  This tension can be quantified by
producing an equivalent ATLAS and CMS combination not from the Large
Observable Set but from the Medium Observable Set giving a relatively
bad $\chi^2/\mathrm{ndf}$ of $10.9/6$.

\begin{figure*}[t]
  \begin{center}
    \subfigure[ ]{
      \centering\includegraphics[width=0.45\textwidth]{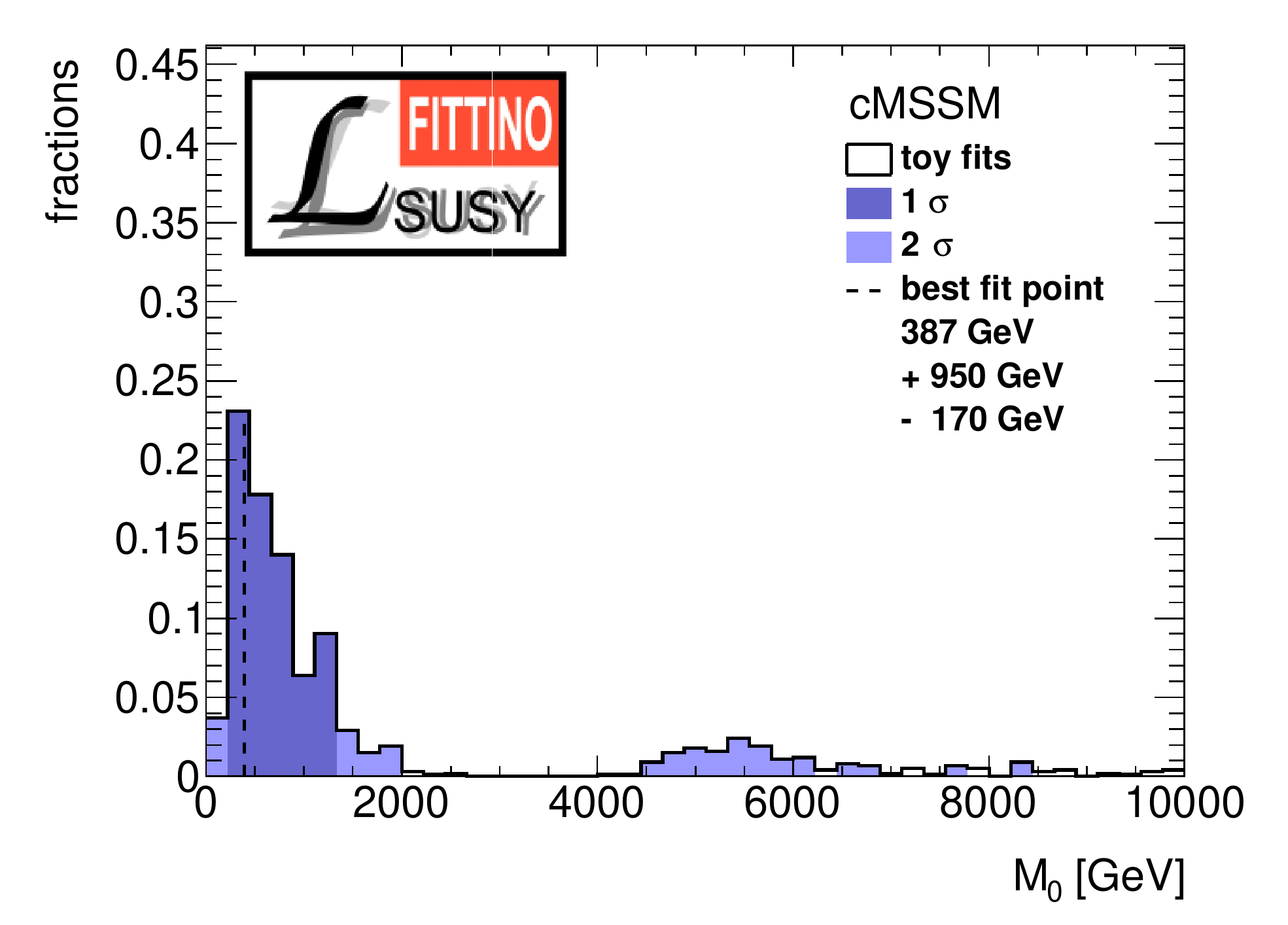}
      \label{fig:MediumObsSet:Toys:M0}  
    }
    \subfigure[ ]{
      \centering\includegraphics[width=0.45\textwidth]{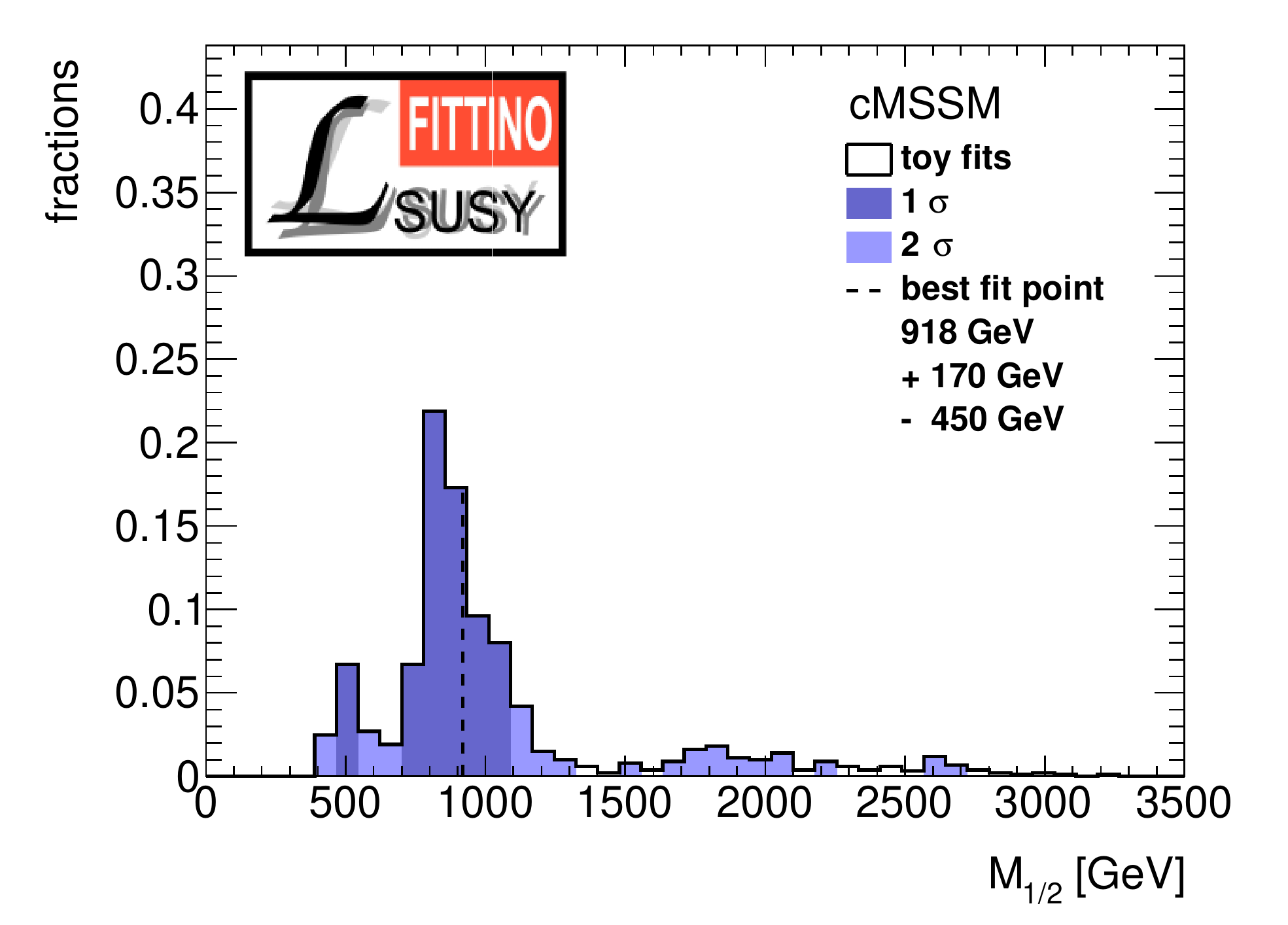}
      \label{fig:MediumObsSet:Toys:M12}  
    }
  \end{center}
  \caption{Distribution of the pseudo best fit values for 
    \subref{fig:MediumObsSet:Toys:M0} $M_{0}$ and \subref{fig:MediumObsSet:Toys:M12} $M_{1/2}$.}\label{fig:MediumObsSet:Toys:Parameters}
\end{figure*}

Finally, as discussed briefly in Section~\ref{sec:methods}, we employ
the Medium Observable Set again to compare the preferred parts of the
parameter space according to the profile likelihood technique
(Fig.~\ref{fig:MediumObsSet:PL:M0M12}) with the parameter ranges
that are preferred according to our pseudo fits. In
Fig.~\ref{fig:MediumObsSet:Toys:M0} and
\ref{fig:MediumObsSet:Toys:M12} we show the 1-dimensional
distributions of the pseudo best fit values for $M_{0}$ and
$M_{1/2}$. The $68\%$ and $95\%$~CL regions according to the total
pseudo best fit $\chi^{2}$ are shown. As expected by the non-Gaussian
behaviour of our fit, some differences between the results obtained by
the profile likelihood technique and the pseudo fit results can be
observed. For the pseudo fits, in both parameters $M_{0}$ and
$M_{1/2}$, the $95\%$~CL range is slightly smaller compared to the
allowed range according to the profile likelihood.  Considering the
fits of the pseudo datasets, $M_{0}$ is limited to values $< 8.5$ TeV
and $M_{1/2}$ is limited to values $ < 2.7$ TeV, while the profile
likelihood technique yields upper limits of $10$ TeV and $3.5$ TeV,
roughly. The differences are relatively small compared to the size of
the preferred parameter space, and may well be an effect of the
limited number of pseudo datasets that have been considered; the use
of the profile likelihood technique for the derivation of the
preferred parameter space can therefore be considered to give reliable
results. However, as discussed above, in order to get an accurate
estimate for the $95\%$-CL regions, the $p$-value would have to be
evaluated at every single point in the parameter space.

%% file: conclusions.tex
\section{Conclusions}\label{sec:conclusions}

In this paper we present what we consider the final analysis of the
cMSSM in light of the LHC~Run\,1 with the program \textsc{Fittino}. 

In previous iterations of such a global analysis of the cMSSM, or any
other more general SUSY model, the focus was set on finding
regions in parameter space that globally agree with a certain set of
measurements, either using frequentist or bayesian
statistics. However, these analyses show that a
constrained model such as the cMSSM has become rather trivial: because of 
the decoupling behaviour at sufficiently high SUSY mass scales
the phenomenology resembles that of the SM with dark
matter. This does however not
disqualify the cMSSM as a valid model of Nature.  In addition, there are
undeniable fine-tuning challenges, but also these do not statistically
disqualify the model. Therefore, before abandoning the
cMSSM, we apply one crucial test, which it has not been performed
before: we calculate the $p$-value of the cMSSM through toy tests.

A likelihood ratio test of the cMSSM against the SM would be meaningless, since
the SM cannot acommodate dark
matter. Thus we apply a goodness-of-fit test of the
cMSSM. As in every likelihood test (also in likelihood ratio tests),
the sensitivity of the test towards the validity of the model
depends on the number of degrees of freedom in the observable set,
while the sensitivity towards the preferred parameter range does not. Thus,
when calculating the $p$-value of the cMSSM, it is
important that the observable set is chosen  carefully. First, only
such observables should be considered for which the cMSSM predictions are, in 
principle, sensitive to the choice of the model parameters, independent of the 
actually measured values of the observables. 
This excludes e.g.{} many LEP/SLD precision observables, for
which the cMSSM by construction always predicts the SM value for
any parameter value. Second, it is important that observables
are combined whenever the corresponding cMSSM predictions are not
independent. Otherwise
the resulting $p$-value would be too large by construction. It should
be noted that the allowed parameter space for all observable sets
studied here is virtually identical. It is only the impact on the
$p$-value which varies.

In order to study this dependence, several observable sets are
studied. The main challenge arises from the Higgs rate
measurements. Since the cMSSM Higgs rate predictions are, in
principle, very sensitive to the choice of model parameters, the
corresponding  
measurements have to be included in a global fit.  
Using the preferred observable sets ``combined'' and ``medium'' (as
described in Section~\ref{sec:results:toy} and
Tab.~\ref{tab:pv:pvalues}), we calculate a $p$-value of the cMSSM of 
4.9\% and 8.4\%, respectively.
In addition, the
cMSSM is excluded at the 98.7\%\,CL if Higgs rate measurements are
omitted.
The main reason for these low $p$-values is the
tension between the direct sparticle search limits from the LHC  and the measured
value of the muon anomalous magnetic moment $(g-2)_{\mu}$. If
e.g. $(g-2)_{\mu}$ is removed from the fit, the $p$-value of the cMSSM
increases to about 50\%. However, there is no justification
for arbitrarily removing one variable \textit{a posteriori}.
 On the other hand, the observable sets could be artificially
chosen to be too detailed, such that there are many measurements for which 
the model predictions cannot be varied individually. This is
the case for the Large Observable Set of Higgs rate observables in
Tab.~\ref{tab:pv:pvalues}, the inclusion of which does thus not represent a
methodologically stringent test of the $p$-value of the cMSSM.

Thus, the main result of this analysis is that the cMSSM is excluded at least at the 
90\% CL for reasonable choices of the observable set.

The best-fit point is in the $\tilde{\tau}$-coannihilation region at
$M_0\approx 500$\,GeV, with a secondary minimum in the focus-point
region at $M_{1/2},M_0\gg2$\,TeV. A comparison of the $p$-values of
coannihilation and focus-point regions can serve as an estimate of a 
likelihood-ratio test between a cMSSM at $M_0\approx 500$\,GeV which can be tested at
the LHC, and a ``SM
with dark matter'' with squark and gluino masses beyond about 5\,TeV. Since the focus point 
manifests a more linear
relation between observables and input parameters in the toy fits, and thus
a more $\chi^2$-distribution like behaviour, it reaches a slightly higher
$p$-value than the $\tilde{\tau}$-coannihilation region. This shows that
even the best-fit region offers no statistically relevant
advantage over the ``SM with dark matter''. Thus, we can conclude that
the cMSSM is not only excluded at the 90\% to 95\%\,CL, but that
it is also statistically mostly indistinguishable from a
hypothetical SM with dark matter.

In addition to this main result, we apply the first complete scan of
the possibility of the existence of charge or colour-breaking minima
within a global fit of the cMSSM. In addition, we calculate the
lifetime of the best fit points. We find that the
focus-point best-fit-region is stable, while the
$\tilde{\tau}$-coannihilation best-fit region is either stable or
metastable, with a lifetime significantly longer than the age of the
Universe.

It is important to note  that the exclusion of the cMSSM
at the 90\%\,CL or more  does in
general not apply to less restricted SUSY
models. The combination of measurements
requiring low slepton and gaugino mass scales, such as $(g-2)_{\mu}$,
and the high mass scales preferred by the SM-like Higgs and the
non-observation of coloured sparticles at the LHC puts the cMSSM
under extreme tension. In the cMSSM these mass scales are
connected. A more general SUSY model where
these scales are decoupled, and preferably also with a complete
decoupling of the third generation sleptons and squarks from the first
and second generation, would easily circumvent this tension.

Therefore, the future of SUSY searches at the LHC should emphasize the
coverage of any phenomenological scenario which allows sleptons, and
preferably also third generation squarks, to remain light, while the
other sparticles can become heavy. Many loopholes with light SUSY
states still exist, as analyses as in~\cite{CMS-PAS-SUS-13-020} show,
and there exist potentially promising experimental anomalies which
could be explained by more general SUSY
models~\cite{Khachatryan:2015lwa,Aad:2015wqa,ATLAS-CONF-2014-033}.

On the other hand, the analysis presented here shows that SUSY does
not directly point towards a non-SM-like light Higgs boson. The
uncertainty on the predictions of ratios of partial decay widths and
other observables at the LHC are significantly smaller than the direct
uncertainty of the LHC Higgs rate measurements. This is because of the
high SUSY mass scale, also for third generation squarks, imposed by
the combination of the cMSSM and the direct SUSY particle search
limits. These do not allow the model to vary the light Higgs boson
properties sufficiently to make use of the experimental uncertainty in
the Higgs rate measurements. This might change for a more general SUSY
model, but there is no direct hint in this direction. The predicted
level of deviation of the light Higgs boson properties from the SM
prediction at the ${\cal O}(1\%)$ level is not accessible even at a
high-luminosity LHC and requires an $e^+e^-$ collider.

In summary, we find that the undeniable freedom in choosing the
observable set -- before looking at the experimental values of the
results -- introduces a remaining softness into the exclusion of the
cMSSM. Therefore, while we might have preferred to find SUSY
experimentally, we find that at least we can almost complete the
second most revered task of a physics measurement: with the
combination of astrophysical, precision collider and energy frontier
measurements in a global frequentist analysis we (softly) kill the
cMSSM.

%% file: acknowledgements.tex
\begin{acknowledgements}
  We thank Sven Heinemeyer and Thomas Hahn for very helpful
  discussions during the preparations of the Higgs boson decay rate
  calculations. This work was supported by the Deutsche
  Forschungsgemeinschaft through the research grant HA 7178/1-1, by
  the U.S. Department of Energy grant number DE-FG02-04ER41286, by the
  BMBF Theorieverbund and the BMBF-FSP~101 and in part by the
  Helmholtz Alliance ``Physics at the Terascale''. T.S.{} is supported
  in part by a Feodor-Lynen research fellowship sponsored by the
  Alexander von Humboldt Foundation. We also thank the Helmholtz
  Alliance and DESY for providing Computing Infrastructure at the
  National Analysis Facility.
\end{acknowledgements}